\documentclass[aps,twocolumn,floatfix,amsmath,amssymb,10pt]{revtex4-2}
\usepackage{graphicx}
\usepackage{dcolumn}
\usepackage{bm}
\usepackage{graphics}
\usepackage{epsfig,color}
\usepackage{array}
\usepackage{dcolumn}
\usepackage{bm}
\usepackage{amsmath,amsfonts,amssymb,amsthm,verbatim}
\usepackage{pstricks}
\usepackage[ansinew]{inputenc}
\usepackage{color}
\usepackage{epsfig}

\newcommand{\PHDG}{^{\vphantom{\dagger}}}

\newcommand{\be}{\begin{equation}}
\newcommand{\ee}{\end{equation}}
\newcommand{\bea}{\begin{eqnarray}}
\newcommand{\eea}{\end{eqnarray}}

\newcommand{\bi}{\begin{itemize}}
\newcommand{\ei}{\end{itemize}}
\newcommand{\bc}{\begin{center}}
\newcommand{\ec}{\end{center}}

\newcommand{\up}{\uparrow}
\newcommand{\down}{\downarrow}

\newcommand{\s}{\sigma}
\newcommand{\reff}{\text{eff}}

\makeatletter

\@ifundefined{textcolor}{}
{%
	\definecolor{BLACK}{gray}{0}
	\definecolor{WHITE}{gray}{1}
	\definecolor{RED}{rgb}{1,0,0}
	\definecolor{GREEN}{rgb}{0,0.5,0}
	\definecolor{BLUE}{rgb}{0,0,1}
	\definecolor{CYAN}{cmyk}{1,0,0,0}
	\definecolor{MAGENTA}{cmyk}{0,1,0,0}
	\definecolor{YELLOW}{cmyk}{0,0,1,0}
}

 \let\old@rule\@rule
 \def\@rule[#1]#2#3{\textcolor{RED}{\old@rule[#1]{#2}{#3}}}

\makeatother

\newlength{\wdth}

\definecolor{rulecolor}{named}{RED}

\begin{document}

\title{Repulsion driven metallic phase in the ground state of the\\ 
half-filled $t-t^{\prime}$ ionic Hubbard chain }

\author{ Gerardo L. Rossini$^{a}$ and George I. Japaridze$^{b,c}$}
\vspace{3mm}

\affiliation{$^{a}$ IFLYSIB-CONICET and Departamento de F\'{\i}sica,
Universidad Nacional de La Plata, 1900 La Plata, Argentina}

\affiliation{$^{b}$ 
Center of Condensed Matter Physics and Quantum Computations,
             Ilia State University, Cholokashvili Avenue 3-5, 0162 Tbilisi, Georgia}

\affiliation{$^{c}$Andronikashvili Institute of Physics,
            Tamarashvili str.~6, 0177 Tbilisi, Georgia}

\begin{abstract}

An unusual metallic phase is argued to develop in the one dimensional ionic Hubbard model, 
at half-filling and zero magnetization, at intermediate electron-electron repulsion $U$
when second neighbors hopping is allowed and tuned close to a topological Lifshitz transition 
(connected with a change of the Fermi surface in the non-interacting system).
The metallic state lies between a band insulator phase at low repulsion and a correlated 
(Mott-like) insulator phase at high repulsion. 
In approaching the later, the model supports 
short range antiferromagnetic order
and spontaneous dimerization of both
bond charge 
and 
nearest neighbors antiferromagnetic correlations. 
A combination of mean field and effective field theory (bosonization) provides an 
analytical understanding of the physical processes underlying the argued phase transitions.
The ground and low energy excited states of finite length chains
are explored by density-matrix renormalization-group (DMRG) calculations, 
providing numerical evidence for the intermediate gapless phase.  
Such finite systems are attainable by cold atoms in  optical lattices for a wide range of the parameter $U$.

\end{abstract}


\maketitle
\newpage
\section{Introduction\label{sec:Introduction}}

Atomic gases stored in artificially engineered optical lattices offer an unique possibility to simulate and study  
condensed-matter systems  with unconventional, or less achievable in actual materials, properties
~\cite{Lewenstein2012-ws,Esslinger_2010,UCA_QS_Hubbard_18}. 
Among the advantages of these systems is the possibility to manipulate 
the strength of the interaction using the Feshbach resonance~\cite{UCA_Feshbach_Rev}, 
what enables to monitor the evolution of the ground state properties of the quantum many-body system with interaction, 
starting from the weak coupling till the limit of very strong interaction. 
Competition between the (kinetic) delocalization energy and interaction is profoundly seen in low-dimensional quantum systems and 
leads to a very rich set of many--body phases displayed in the remarkable ground state (GS) properties 
of these systems~\cite{GNT_Book,Giamarchi_Book}.

Optical lattices can be generated in various geometries, including two dimensional
 triangular~\cite{UCA_triangle_2010,Triangular_lattice_2019}, Kagome~\cite{UCA_Kagome_Lattice_2012},
 hexagonal~\cite{Tarruel_12,Esslinger_13} structures 
 as well as quasi-one-dimensional few chain systems with zig-zag~\cite{Zig-Zag_ladder_2016} or 
 ladder~\cite{Kang_etal_18,Kang_etal_20} geometry.  
 In addition, the optical engineering allows to manipulate the details of lattice structure, 
 in particular to introduce a bias for atom occupation energy on neighboring sites and thus to create 
 a bipartite lattice~\cite{Hemmerich_11a,Hemmerich_11b}  or ladder with non-equivalent legs ~\cite{Esslinger_etal_2015}.   
 This makes the ground state phase diagram of the system even more complex and 
 opens the possibility to experimentally investigate the nature of various quantum phase transitions 
 between different phases with remarkable properties.  In particular, fermionic atomic gases with repulsion on optical 
 lattices provide an excellent testing ground to study 
 {\em insulator-insulator} and {\em metal-insulator} transitions driven by the interplay between the effects caused by correlations,
  geometrical frustration and non equivalence of atomic sub-lattices~\cite{UCA_QS_Hubbard_08a, Aligia_22, Walter_23}.  
  Also emergent new effects connected with  the topological Lifshitz transition~\cite{Volovik_17} are 
  of the prime current interest~\cite{Ruchman_Altmann_17,Mazza_Simon_Roux_21,Aditya_Sen_21,Cazalila-etal_22,Halati_Giamarchi_22,Beradze_Nersesyan_22}. 
 
In this paper we consider the one-dimensional model of interacting fermions given by the following Hamiltonian 
\begin{align}
 \label{eq:t1t2_IH_model}
 \begin{split}
  {\cal H} =&  -t\sum_{i,\sigma}^{L} \left(c^{\dagger}_{i,\sigma} c\PHDG_{i+1,\sigma}+ \mathrm{H.c.}\right) \\
  &+t^{\prime}\sum_{i,\sigma}^{L} \left(c^{\dagger}_{i,\sigma} c\PHDG_{i+2,\sigma}+ \mathrm{H.c.}\right)\\
  & +\frac{\Delta}{2} \sum_{i,\sigma}^{L}(-1)^{i}n_{i,\sigma}+U\sum_{i}n_{i,\up}n_{i,\down}\,.
 \end{split}
\end{align}
Here $c^{\dagger}_{i, \sigma}$ $(c\PHDG_{i,\sigma})$ creates (annihilates) a fermion with
spin ${\sigma = \up,\down}$  on site $i$ and ${n_{i,\sigma}=c^{\dagger}_{i,\sigma}c\PHDG_{i,\sigma}}$ is the spin $\sigma$ 
particle density operator. 
The nearest neighbor (n.n.) hopping amplitude is denoted by $t$, 
the next-to-nearest neighbor (n.n.n.) hopping amplitude by $t^\prime$
(${t,t^\prime>0}$), $\Delta$ is the potential energy difference between neighboring sites, ${U}$ is the on-site Coulomb repulsion. 
The Hamiltonian~\eqref{eq:t1t2_IH_model} commutes with the number operator of particles with spin $\sigma$, 
${\cal N}_{\sigma}=\sum_{i}n_{i,\sigma}$. 
Below in this paper we restrict our consideration to the case of half-filled band with zero magnetization, 
with particle number eigenvalues ${N_\uparrow=N_\downarrow=L/2}$, and to repulsive interaction ${U>0}$.

\begin{figure}
	\begin{centering}
		\includegraphics[scale=0.6]{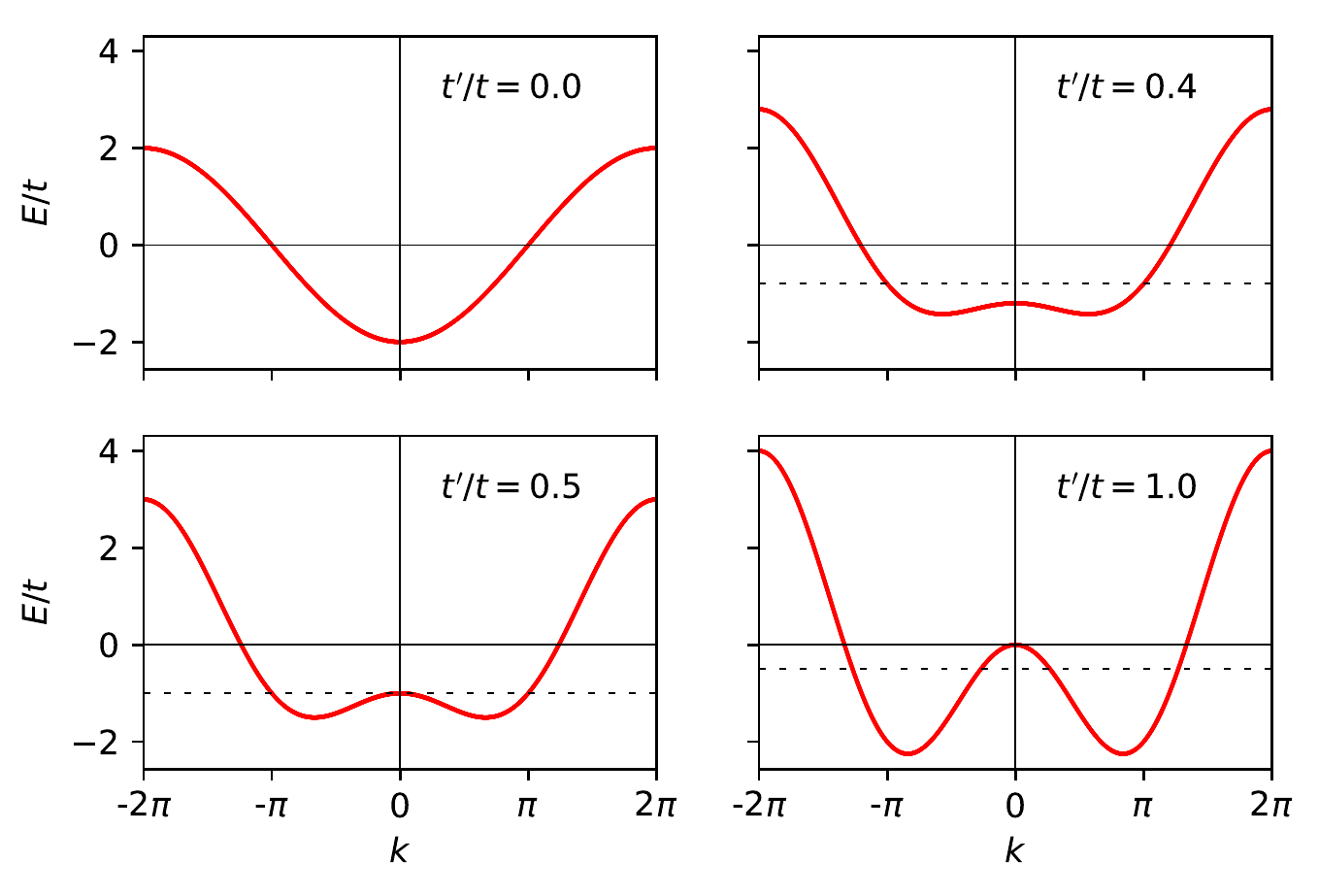}
\caption{Dispersion relation $E_k$
	for the ${t-t^{\prime}}$ (non-ionic, $\Delta=0$) chain at different values of ${t^{\prime}}/t$.
	The dashed line indicates the chemical potential at half-filling.  At $t^{\prime}=0.5\,t$ a Lifshitz transition takes place, changing the structure of the Fermi surface from two to four Fermi points.
		}
		\label{fig: t1-t2 energy}
	\end{centering}
\end{figure}

For ${\Delta=0}$, the Hamiltonian corresponds to the ${t-t^\prime}$ Hubbard model~\cite{MH_95,Fabrizio_96}
in the case of half-filled band, the prototype model to study the metal-insulator transition in 
1D~\cite{Kuroki_97,DaulNoack_00,AebBaerNoack_01,Torio_03,Japaridze_etal_07a,Satoshi_08}. 
At ${t^\prime<0.5t}$ the system is in a gapped insulating phase for arbitrary ${U>0}$, 
but at ${t^\prime>0.5t}$ is characterized by the quantum phase transition from a
charge gapless  metallic behavior at ${U<U_{c}}$ into an insulating phase at ${U>U_{c}}$ ~\cite{Kuroki_97,DaulNoack_00,AebBaerNoack_01}. 
The qualitative change of the ground state properties of the system at ${t^{\prime}>0.5t}$  
emerges as the result of the topological Lifshitz transition in the ground state of free system, 
where the number of Fermi points doubles~\cite{Fabrizio_96} (see Fig.\ \ref{fig: t1-t2 energy}). 
It has been shown that at fixed $U$ and increasing $t^\prime$ the insulator to metal transition 
is described in terms of the commensurate-incommensurate transition \cite{JN_78,PT_79} with
a transition curve determined by the relation  $M_c(U) =2t_{c}^\prime-t^{2}/t_{c}^{\prime}$, 
where $M_c(U)$ is the charge (Hubbard) gap at the given ${U}$ and ${t^\prime=0}$~\cite{Japaridze_etal_07a}.  

For  ${t^\prime =0}$ and ${\Delta \neq 0}$ the Hamiltonian (\ref{eq:t1t2_IH_model}) 
describes the ionic Hubbard model (IHM) \cite{Hubbard_Torrance_81,Nagaosa_86,Egami_93,Fab_99}. 
At finite ${\Delta}$ the translational invariance is explicitly broken, the lattice unit is doubled, 
and the density imbalance between neighboring sites shows up via the presence of 
long-range-ordered (LRO) charge density wave (CDW) pattern in the ground state for arbitrary ${U>0}$ \cite{Brune_etal_03}. 
On the other hand the repulsive Hubbard interaction  suppresses density inhomogeneities and 
favors antiferromagnetic ordering on neighboring sites.
Competition between these tendencies is resolved in the ground state phase diagram via the presence of two, 
excluding each other phase sectors - the band insulating CDW phase at  ${U<U_{c1}}$ and correlated Mott insulating phases at ${U>U_{c2}}$, 
separated by a narrow intermediate bond-ordered wave (BOW) phase \cite{Fab_99}. 
The nature of the corresponding 
phase transitions has been also first established within the continuum-limit bosonization description,  
showing the Ising type (charge) transition at ${U_{c1}}$ 
from a CDW band insulator phase to a BOW insulator phase and 
the second (spin) Kosterlitz-Thouless type transition, 
at ${U_{c2}}$, from the BOW to a correlated Mott insulator \cite{Fab_99}. 
Subsequent numerical studies have unambiguously proven this phase diagram  \cite{Aligia_01, Brune_etal_03,Furusaki_04,Manmana_04,Tincani_09}.

\begin{figure}
\begin{centering}
\includegraphics[scale=0.6]{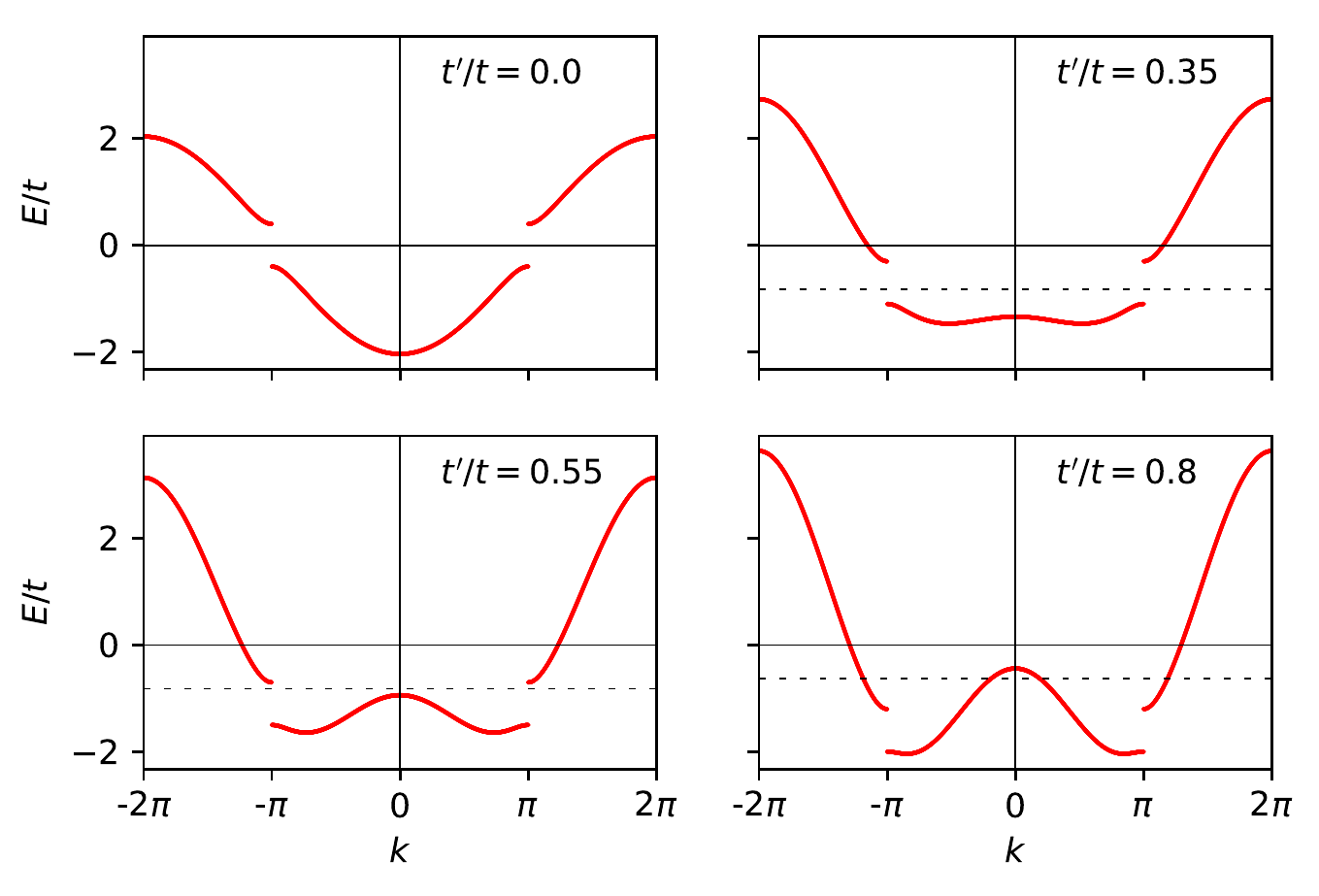}
\caption{
	Dispersion relation $E_k$ for the $t-t^{\prime}$ ionic chain at  ${\Delta=0.8\,t}$ and different values of $t^{\prime}/t$. The dashed line indicates the chemical potential at half-filling.
	The  panels with $t'=0.0, \,0.35\,t$ show a band insulator with direct gap, while the one with $t'=0.55\,t$ illustrates an indirect gap and quadratic dispersion for quasi-particles and holes  close to the insulator-metal transition. 
	In the panel with $t'=0.8\,t$ the system  becomes gapless, with well defined linear dispersion for quasi-particles and holes
	around four Fermi points.
	The unit cell has two sites, here the Brillouin zone is expanded to show the two dispersion branches side by side.
	}  
\label{fig: t1-t2 energy gaps_D}
\end{centering}
\end{figure}

At ${U=0}$ the model can be easily diagonalized in the momentum space (see Appendix \ref{app: A}). 
For ${t^{\prime} < t_{\ast}^{\prime}=0.5t\sqrt{1+(\Delta/2t)^{2}}-\Delta/8} $ (assuming $\Delta>0$) 
the ground state corresponds to the standard CDW band insulator with direct gap; 
at ${t_{\ast}^{\prime} < t^{\prime} < t^{\prime}_{c}}$ to the band insulator (BI) with indirect gap 
and at  ${t^{\prime} > t^{\prime}_{c}=0.5t\sqrt{1+(\Delta/2t)^{2}}+\Delta/8}$ to the metal. 
In this case the Lifshitz transition is shaded by the presence of the band gap 
and displays itself in the insulator-metal transition, where a Fermi surface with four points opens (see Fig.\ \ref{fig: t1-t2 energy gaps_D}). 

Inclusion of the Hubbard repulsion into the scheme introduces an additional set of complexity, 
both the metal and insulating phases experience transition into different insulating phases at strong repulsion. 
In a recent publication this problem has been addressed within the mean-field approximation~\cite{Sekania_etal_22}. 
It has been shown that unconventional insulating phases, 
characterized by a spin and  charge--density modulation with a wavelength equal to four lattice units, 
become energetically favorable above the Lifshitz transition 
and almost completely wipe out the metallic phases. 
This type of density modulations are absolutely natural for the interacting fermions with n.n.n. hopping
and emerge in the system at ${t^\prime \gg t^{\prime}_{c}}$ 
as the result of the opening of four Fermi points and the explicit breaking of translational symmetry by the finite ionic term. 

However, in the direct proximity of the insulator-metal (Lifshitz) transition, 
at ${t_{\ast}^{\prime} < t^{\prime} < t^{\prime}_{c}}$,  
metallic properties of the free system are described by particles and holes with quadratic dispersion 
and thus details of the phase diagram deserve a more accurate analysis than the previous mean-field approximation. 
In this paper we address this problem and find a different scenario.
 An analytical study based on both  an improved  mean-field approximation 
	and tailored bosonization tools allows to understand the underlying 
physical processes responsible for the complex nature of the phase diagram. 

\begin{figure}  
	\begin{centering}
		\includegraphics[scale=0.7]{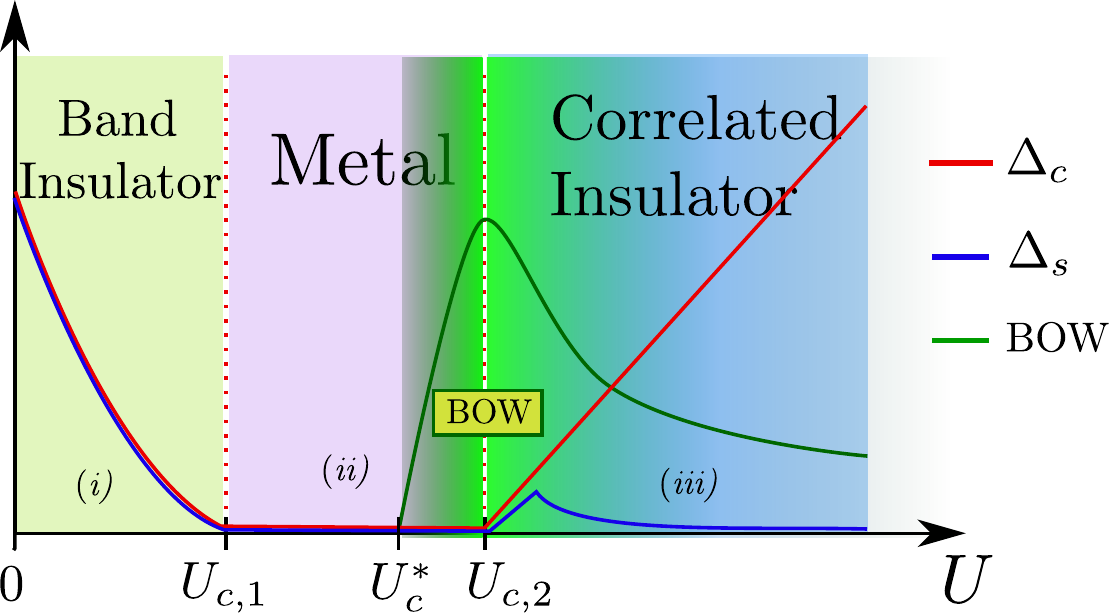}
		\caption{Schematic phase diagram  suggested by our numerical results. 
			Fixed  $t'/t$ is tuned so that the non-interacting system is
		close to the Lifshitz transition, still bearing an indirect excitation gap 
		(see Fig.\ \ref{fig: t1-t2 energy gaps_D}, lower left panel). 
			The Hubbard repulsion $U$ drives the system from a band insulator into an unconventional metal ($U_{c,1}$)
			before reaching the correlated insulator phase  ($U_{c,2}$).
			The charge gap $\Delta_c$ is plotted in red, the spin gap $\Delta_s$ in blue, and the BOW order amplitude in green. 
			Areas in solid colors identify the ground state phase according to the charge gap, 
			while the green gradient indicates the presence of spontaneous BOW order ($U>U^\ast_c$)  starting inside the metallic phase. 
			\label{fig:phase_diagram}
		}
		\end{centering}
	\end{figure}

Density-Matrix Renormalization Group (DMRG) computations,
setting $t$, $t'$ and $\Delta$  where the non-interacting system is gapped but close to the Lifshitz point,  
support the existence of a metallic phase at intermediate Hubbard repulsion $U$. 
 Though at present we are not able to properly extrapolate finite size results into 
a controlled thermodynamic limit, 
our numerical exploration suggests the picture shown in  Fig.\ \ref{fig:phase_diagram}.
In the considered range of parameters, the ground state phase diagram of the 
system as a function of the on-site Hubbard repulsion $U$ consists of three phases: 
At $0 < U < U_{c,1}$ the band insulating CDW phase,  
for $U_{c,1} < U < U_{c,2}$ a repulsion driven metallic phase 
and for $U > U_{c,2}$ a correlated insulator (CI) phase. 
The LRO CDW pattern is clearly present, with decreasing amplitude, in all these phases. 
A spontaneous BOW order appears inside the metallic phase,  with increasing amplitude towards its edge; 
this amplitude starts to decay as soon as the charge gap reopens, however it remains finite 
in the CI phase and continuously evolves into the spin dimerization pattern at $U\rightarrow \infty$. 
 
The paper is organized as follows. 
	In Section \ref{sec:MeanField} we present a mean-field approach leading to a 
	renormalization of the ionicity parameter $\Delta$ due to electron-electron interactions; 
	we explore the appearance of a metallic phase within this regime. 
	In Section \ref{sec:bosonization} we introduce a bosonization scheme allowing 
	to analyze on equal footing the role of ionicity $\Delta$, 
	Hubbard repulsion $U$ and n.n.n. hoping $t'$; 
	within this framework we discuss the different possible ground state phases of the present model. 
	We also identify a parameter region where such phases are realized. 
	In Section \ref{sec:Numerics} we numerically explore the model with the DMRG technique, 
	selecting intermediate $t'$ and $\Delta$ and a full range for the Hubbard repulsion $U$. 
	Finally in Section \ref{sec:Summary} we summarize and discuss the obtained results.

\section{Qualitative estimations}
\subsection{Self-consistent approach \label{sec:MeanField}}

Because the translation symmetry of the system is explicitly broken by the $\Delta$ term in Eq.\ (\ref{eq:t1t2_IH_model}), 
an alternating pattern of charge density is present in the ground state 
at arbitrary $U$ \cite{Brune_etal_03}. 
For further analysis it is convenient to subtract from the density operators their vacuum expectation values 
rewriting them in the following way
\begin{equation}\label{eq:MF_n}
n_{i,\sigma}=\frac{1}{2}\left[1 - (-1)^{i}\delta\rho_{0}(U)\right]+:n_{i,\sigma}:
\end{equation}
where  :$\,$: denote fluctuations on top of the GS value and $\delta\rho_{0}(U)$ 
is the amplitude of the CDW pattern present in the ground state at given $U$. 
Here we take into account that ${\langle\,n_{i,\up}\,\rangle =\langle\, n_{i,\down}\,\rangle}$. 
Using Eq.\ (\ref{eq:MF_n}) the Hamiltonian in Eq.\ (\ref{eq:t1t2_IH_model}) can be rewritten in the following way 
\begin{eqnarray}\label{eq:t1t2_IH_model_1}
 {\cal H} &=&     -t\sum_{i,\sigma}^{L} \left(c^{\dagger}_{i,\sigma} c\PHDG_{i+1,\sigma}+
  \mathrm{H.c.}\right)\nonumber\\  
 &+&t^{\prime}\sum_{i,\sigma}^{L} \left(c^{\dagger}_{i,\sigma} c\PHDG_{i+2,\sigma}+ \mathrm{H.c.}\right)\\
& + &\frac{\Delta_{r}}{2} \sum_{i,\sigma}^{L}(-1)^{i}:n_{i,\sigma}:  +  U\sum_{i}:n_{i,\up}::n_{i,\down}: \, ,\nonumber
\end{eqnarray} 
where
\begin{equation}\label{eq:Delta_r}
\Delta_{r}(U)=\Delta-U\delta\rho_{0}(U)\, .
\end{equation}
Thus even in the gapped band insulating phase, where the charge fluctuations 
are suppressed and at weak-coupling one could ignore their interaction 
in the last term of Eq.\ (\ref{eq:t1t2_IH_model_1}), 
the contribution of the on-site Hubbard term is crucial and 
manifest in the renormalization of the ionic gap given in Eq.\ (\ref{eq:Delta_r}). 


Below in this subsection we restrict our consideration to  the mean-field approximation and neglect the  
scattering of quasiparticles (blocked by the band gap) on top of the Fermi surface 
given by the last term in Eq.\ (\ref{eq:t1t2_IH_model_1}).  
In this case the Hamiltonian can be easily diagonalized in momentum space (see Appendix \ref{app: A}) to give 
\begin{equation}
  {\cal H}_{t-t'-\Delta_r}
  =
  \sum_{k,\s}
    \left(
      E_{k}^{-}  \alpha^\dagger_{k,\sigma}  \alpha\PHDG_{k,\sigma}
      +
      E_{k}^{+} \beta^\dagger_{k,\sigma} \beta\PHDG_{k,\sigma}   
    \right)
  \,, 
\end{equation}
where
\begin{equation}
 \label{eq:Ek1D}
  E_{k}^{\pm}
  =
  \varepsilon^\prime_{k}
  \pm
  \sqrt{\varepsilon_{k}^2 + (\Delta_{r}/2)^2}
  \,
\end{equation}
are the energy dispersions for $\alpha$- and $\beta$-quasiparticles, corresponding to the "lower" and "upper" bands, respectively.

In the ground state of the half-filled system the $L$ lowest energy states are filled and the rest $L$ are empty.
For ${t^{\prime}\leq 0.5t}$, ${E^-_k}$ and ${E^+_k}$  are separated with a direct gap equal to ${\Delta_{r}}$;
all states in the "lower" band are occupied whereas in the "upper" band all states are empty; the system is in the insulating state. For ${t^{\prime} > 0.5t}$, with increasing ${t^{\prime}}$ (or reducing ${\Delta_{r}}$) bands might overlap, 
due to the $k$-dependent energy shift $\varepsilon^\prime_k$, and the system experience a transition into the metallic phase.

At given values of the parameters $t$ and  $t^\prime$, it is useful to introduce a critical value of the effective ionicity parameter $\Delta_r^{cr} \geq 0$ 
\begin{equation}\label{eq:Delta_c}
  \Delta_{r}^{cr} =
  \left\{
    \begin{array}{ccc}
      4 t^\prime - t^{2}/t^\prime & &\text{for }  t^\prime \geqslant 0.5 t, \\[0.5em]
      0                           & &\text{otherwise,  }
    \end{array}
  \right.
\end{equation}
corresponding to the metal-insulator transition:  for ${|\Delta_{r}| > \Delta_{r}^{cr}}$ (${|\Delta_{r}| < \Delta_{r}^{cr}}$), 
the system is in an insulating (metallic) state. Note that for ${t^\prime < 0.5t}$ the system remains in the insulating phase for any finite value of $|\Delta_{r}|$.

Inserting in Eq.\ (\ref{eq:Delta_r}) the analytical expression for the amplitude of the CDW modulation in the insulating phase,
\begin{eqnarray}
\delta\rho_{0} & = & \frac{\Delta_{r}\kappa{K}(\kappa)}{2\pi t}\, \label{eq:delta_rho_0}
\end{eqnarray}
where $K(\kappa)$ is the complete elliptic integral of the first kind 
with the modulus ${\kappa(t,\Delta_{r})=\left[1+\left({\Delta_{r}}/{4t}\right)^{2}\,\right]^{-\frac{1}{2}}}$
(see  Eq.\  (\ref{eq:delta_rho}) in Appendix \ref{app: A}),
 we obtain a self-consistent equation for $\Delta_{r}$ 
\begin{equation}
\Delta_{r}=\Delta-\frac{U\,\Delta_{r}\kappa(t,\Delta_{r})\,K(\kappa(t,\Delta_{r}))}{2\pi t}
\label{eq: self-consistency for Delta_r}
\end{equation}
that can be solved iteratively for given $U$.
%

\begin{figure}
	\begin{centering}
		\includegraphics[scale=0.5]{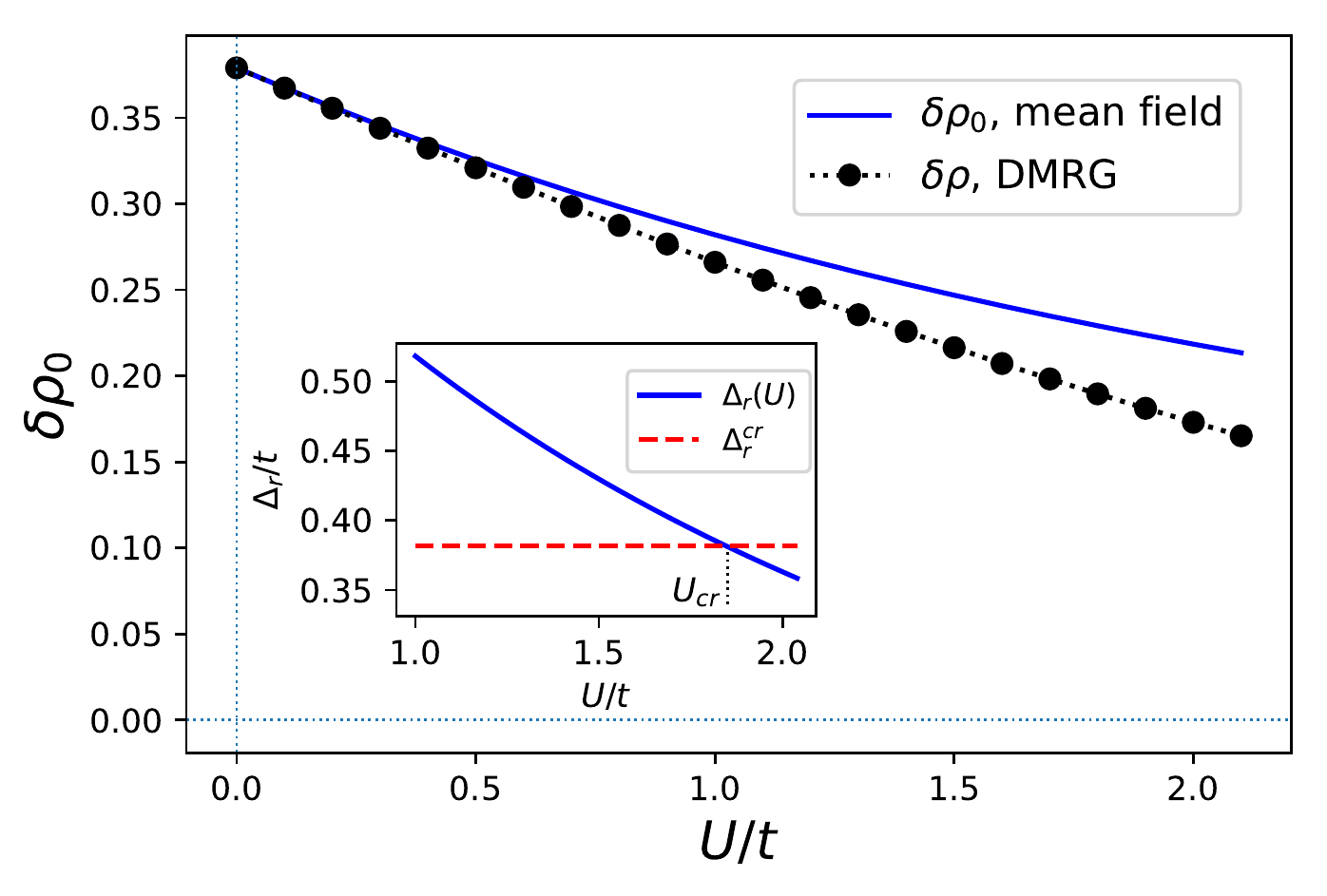}
	\end{centering}
	\caption{
		The self-consistent solution for ${\delta\rho_{0}(U)}$ computed for ${t^{\prime}=0.55t}$, ${\Delta=0.8t}$ is shown in solid blue.  
		Actual DMRG data for ${\delta\rho(U)}$ is shown in black dots for comparison. 
		Inset: self-consistent solution for the renormalized
		$\Delta_{r}$ as a function of $U$ (in solid blue).
		The dashed red line is the critical value for gap closing, ${\Delta_{r}^{cr}=4t^{\prime}-t^{2}/t^{\prime}}$. 
		The intersection occurs at $U_{cr} \approx 1.85\,t$.}
	
	\label{fig:mean_field-solution}
	
\end{figure}
%

The results, for  $t^{\prime}=0.55t $ and $\Delta=0.8t$, are shown in Fig.\ \ref{fig:mean_field-solution}.
In the inset we first show the  renormalized $\Delta_{r}$ as a function of $U$: 
one can see that $U$ competes with the bare ionicity $\Delta$, reducing $\Delta_{r}$. 
Eventually, provided $t^{\prime}>0.5t$, the band gap closes at a critical point $U_{cr}$ 
when 
\begin{equation}
\Delta_{r}(U_{cr})=\Delta_{r}^{cr} \equiv 4t^{\prime}-t^{2}/t^{\prime} \, ,
\end{equation}
driving the system into a metallic phase. 
For the given parameters this occurs at  $U_{cr} \approx 1.85\,t$, 
in qualitative agreement with $U_{c,1} \approx 2.2 \,t$  suggested by the DMRG data discussed in Sec.\ \ref{subsec:Energy gaps}.

Once having ${\Delta_{r}(U)}$ one can compute the CDW amplitude ${\delta\rho_{0}(U)}$ from Eq.\ (\ref{eq:delta_rho_0}),
which is shown in the main panel of Fig.\ \ref{fig:mean_field-solution} in good agreement with exact DMRG data discussed in  Sec.\ \ref{subsubsec:CDW}. 
Notice that, as the  renormalized $\Delta_{r}$ decreases, the  CDW amplitude is also reduced by electron repulsion $U$.

\subsection{Bosonization approach\label{sec:bosonization}}

In this subsection we use the bosonization technique to obtain a qualitative description 
of the low-energy properties of the Hamiltonian in Eq.\ (\ref{eq:t1t2_IH_model_1}). 
We restrict our consideration to  the  weak-coupling case ${\Delta_{r}, U \ll t}$ and ${t^{\prime}\simeq 0.5t}$,  
 \emph{i.e.\ }the close proximity to the insulator-metal transition. 

Because for the selected set of model parameters the spectrum of the free system is either gapped or, 
in a metallic phase, has a quadratic dispersion, the 
straightforward application of the bosonization technique is not possible. 
Therefore, we follow the route developed earlier in studies of the standard IHM~{\cite{Fab_99}},  
where one starts the description from the weak-coupling case, 
linearizes the spectrum in the vicinity of the two Fermi points $k_{F}=\pm \pi/2$  (2-FP approach)  and 
goes to the continuum limit by the substitution
\be
c_{n\sigma} \rightarrow   i^{n} R_{\s}(x) +
(-i)^n L_{\s}(x)\, , \label{linearization}
\ee
where $R_{\s}(x)$ and $L_{\s}(x)$ describe right-moving and
left-moving fermionic particles, respectively. 
This approach allows to treat, within the effective continuum-limit description,  
the gap "creating" ($\Delta_{r}$ and $U$) and gap "destructing" ($t^{\prime}$) terms on an equal footing 
and thus in a transparent way display the character of their competition~{\cite{Japaridze_etal_07b}}. 

Within the framework of 2-FP approach 
the ionic (${\Delta_{r}}$) and the Hubbard (${U}$) terms appear as the scattering processes responsible for generation of a gap in the excitation spectrum. 
The staggered ionic potential introduces a single particle backward scattering process 
${H_{\Delta_r} \sim \Delta_{r}\int dx \sum_{\sigma} (R^{\dagger}_{\s}L^{\phantom{\dagger}}_{\s}+ h.c.)}$ 
and is responsible for generation of equal excitation gaps in each spin subsystem 
\emph{i.e.\ }for formation of the BI phase. The repulsive Hubbard term, via the correlated umklapp scattering processes
${H_{umk} \sim U\int dx 
	(R^{\dagger}_{\up}R^{\dagger}_{\downarrow}
	L^{\phantom{\dagger}}_{\downarrow}L^{\phantom{\dagger}}_{\uparrow}+ h.c.)}$,
is responsible for the formation of the correlated Mott gap in the charge excitation spectrum.  

Development of the gap in the excitation spectrum stabilizes the corresponding band and correlated insulating phases, respectively. 
However since the elementary excitations in the BI and Mott insulating phases are topologically distinct,  
they expel each other and  at $t^{\prime}=0$, in the ground state of the half-filled IHM the BI and Mott insulating phases are separated by the intermediate 
BOW phase~\cite{Fab_99}.

Note that both of the above discussed scattering processes are intimately connected with the selected structure of the Fermi surface 
with two Fermi points ${\pm \pi/2}$ separated by ${\pi}$ and become incommensurate at any change of this condition. 
The "gap destructing" effect of the ${t^{\prime}}$ term is directly connected with a change of the commensurate structure of the Fermi surface. 
To maintain the half-filling and therefore to incorporate accurately the effect of ${t^{\prime}}$-term within the 
used 2-FP approach, 
one has to compensate the shift of the Fermi energy ${\delta E_F}$ introduced by the ${t^{\prime}}$ term  
by a corresponding change of the chemical potential term ${\delta\mu({\cal N}_{\up}+{\cal N}_{\down})}$, where
\be
\delta\mu=-\delta E_F = \left \{
\begin{array}{cl}
2 t^\prime, & t^\prime < 0.5t \\
t^2/2t^{\prime}, & t^\prime > 0.5t \;  \\
\end{array}
\right .\, 
\ee
and {${\cal N}_{\up}+{\cal N}_{\down}$} is the total number of electrons operator. 
Now using the substitution (\ref{linearization}) to express 
the {n.n.n.} hopping in term of right and left fields, we obtain
${H_{t^{\prime}}= -2t^{\prime}\int dx \sum_{\sigma} 
\left(R^{\dagger}_{\s}R^{\phantom{\dagger}}_{\s}+
L^{\dagger}_{\s}L^{\phantom{\dagger}}_{\s}\right)=-2t^{\prime}({\cal N}_{\up}+{\cal N}_{\down})}$. 
and thus the total contribution of the {n.n.n.} hopping term into the effective field theory is given by the 
chemical potential term ${\mu_{\text{eff}}({\cal N}_{\up}+{\cal N}_{\down})}$, with~{\cite{Japaridze_etal_07b}}
\be
\mu_{\text{eff}} =  \left\{
\begin{array}{lll}
	0  & \mbox{for} & t^{\prime} < 0.5t \\
	2t^{\prime} -{t^2}/{2t^{\prime}} \neq 0  & \mbox{for}  &	t^{\prime} > 0.5t\;  
\end{array}
\right .\, .
\label{ChemPot_eff}
\ee

The right and left components of the Fermi fields  
can be bosonized in a standard way 
\begin{eqnarray}\label{Psi_RL_Bosonization}
R_{\sigma}(x) & \rightarrow & \frac{1}{\sqrt{2\pi\alpha_{0}}}
e^{ i\sqrt{4\pi}\phi_{R\sigma}(x)}\nonumber\\
\\
L_{\sigma}(x) & \rightarrow & \frac{1}{\sqrt{2\pi\alpha_{0}}}
e^{- i\sqrt{4\pi}\phi_{L\sigma}(x)}\, \nonumber,
\end{eqnarray}
where $\phi_{R\sigma}$ ($\phi_{L\sigma}$) are right(left)-moving
Bose fields and $\alpha_{0}$ is an infrared cutoff.
We define the conjugate fields $\phi_\sigma=\phi_{R\sigma} + \phi_{L\sigma}$ and  
$\theta_\sigma=\phi_{L\sigma} - \phi_{R\sigma}$, 
which possess commutation relations ${[\phi_{\s}(x),\theta_{\s}(x^{\prime})]}=i\pi\delta(x-x^{\prime})$.
We define the charge
\begin{equation}
\phi_{c} = {\textstyle \frac{1}{\sqrt{2}}} (\phi_{\uparrow} +
\phi_{\downarrow}), 
\qquad 
\theta_{c} = {\textstyle
\frac{1}{\sqrt{2}}} (\theta_{\uparrow} + \theta_{\downarrow })
\label{bos_carge}
\end{equation}
and spin fields
\begin{equation}
\phi_{s} = {\textstyle \frac{1}{\sqrt{2}}} (\phi_{\uparrow} -
\phi_{\downarrow}),\qquad \theta_{s} = {\textstyle
\frac{1}{\sqrt{2}}} (\theta_{\uparrow} - \theta_{\downarrow})
\label{bos_spin}
\end{equation}
to describe corresponding degrees of freedom. After some standard algebra  \cite{Giamarchi_Book}
and a rescaling of the fields we arrive at the following bosonized
version of the Hamiltonian (\ref{eq:t1t2_IH_model_1}):
\bea\label{Hamilt_v1}
{\cal H} =  \int \,dx \big[\,h_{s} + h_{c} + h_{cs}\,\big]\, ,
\eea
where
\bea
&&\hspace{-8mm}  h_{s} =   \frac{v_{s}}{2}\big[(\partial_{x}\phi_{s})^2
+(\partial_x \theta_{s})^2\big]
+ \frac{m^{0}_{s}}{2 \pi^{2} a_{0}^{2}}\cos\sqrt{8\pi}\phi_{s},
\label{SGs}\\
&&\hspace{-8mm}h_{c}  = \frac{v_{c}}{2}\big[(\partial_{x}\phi_{c})^2
+ (\partial_x \theta_{c})^2\big]- \frac{m^{0}_{c}}{2 \pi^{2} a_{0}^{2}}\cos\sqrt{8\pi K_{c}}\phi_{c}\nonumber \\
&&\hspace{10mm}  - \mu_{\reff}\sqrt{\frac{K_{c}}{2\pi}}\partial_{x}\phi_{c}\, , 
\label{SGc}\\
&&h_{cs}  = - \frac{\Delta_{r}}{\pi a_{0}}\sin\sqrt{2\pi K_{c}} \phi_{c} \cos\sqrt{2\pi}\phi_{s}\, . \label{H_CS_BOS} 
\eea
Here  $m^{0}_{s}\sim U$ and $m^{0}_{c} \sim U$ are the bare values of coupling constants, 
the charge stiffness parameter is ${K_{c} <1}$ at $U>0$  and ${v_{s}}$ and ${v_{c}}$ are velocities of spin and charge excitations. 

At ${\Delta_{r}=0}$ the Hamiltonian (\ref{Hamilt_v1}) describes the Mott-insulator--metal transition in the ground state 
of the half-filled Hubbard chain, 
caused by the change of chemical potential $\mu_{\reff}$~\cite{Japaridze_etal_07a}. 
Respectively, at $m^{0}_{s}=m^{0}_{c}=0$,  the BI--metal transition  in the ground state of the
n.n. free ionic chain
(see for details Appendix \ref{app: B}). 
In each of these limiting cases the model reduces to the standard Hamiltonian of the sine-Gordon model with topological term, 
describing the commensurate-incommensurate transition~\cite{JN_78,PT_79}, 
which has been intensively studied in the past using bosonization and the Bethe ansatz \cite{JN_79,JNW_1984}. 
In each case, the transition into the metallic phase takes place when the chemical potential exceeds the corresponding charge gap. 

In the considered case of coupled fields with two separate sources for the charge gap formation the situation is more complicate. 
To step forward let us first eliminate the chemical potential term by the gauge transformation
\bea
\sqrt{2\pi}\phi_{c}(x) \rightarrow \sqrt{2\pi}\phi_{c}(x) +  \frac{\mu_{\reff}\sqrt{K_{c}}}{ v_c }x
\eea
and rewrite the Hamiltonian density in (\ref{Hamilt_v1}) in the following form
\bea
&&\hspace{-8mm} h_{s} =   \frac{v_{s}}{2}\big[(\partial_{x}\phi_{s})^2
+(\partial_x \theta_{s})^2\big]+\frac{m^0_s}{2 \pi^{2} a_{0}^{2}}\cos\sqrt{8\pi}\phi_{s},\label{H_S_BOS} \\
&&\hspace{-4mm} h_{c}  = \frac{v_{c}}{2}\big[(\partial_{x}\phi_{c})^2+ (\partial_x \theta_{c})^2\big]\nonumber \\
&&\hspace{+2mm} -\frac{m^0_c}{2 \pi^{2} a_{0}^{2}}\cos(\sqrt{8\pi K_{c}}\phi_{c}+ 2x/l_{\mu})\, ,\label{H_C_BOS} \\
&&\hspace{-8mm}h_{cs}=
 -\frac{\Delta_{r}}{\pi a_{0}}\sin\left(\sqrt{2\pi K_{c}}\phi_{c}+x/l_{\mu}\right) \cos\sqrt{2\pi}\phi_{s},\label{H_CS_BOS} 
\eea
where, the characteristic length
\bea
l_{\mu}& = & \frac{ v_c }{\mu_{\reff}\sqrt{K_{c}}}
\eea
determines the distance, above which the effects of "doping" (\emph{i.e.\ }deviation of the Fermi points from 
$\pm\pi/2$) become visible. On the other hand, each of the gap generating terms separately can be characterized by 
its own length-scales  $l_{\Delta} \sim  v_{F}/\Delta_{r}$ -- the ionic term -- and $l_{M_{c}} \sim v_{F}/M_{c}$   
-- the Hubbard term -- where $M_{c}$ is the correlated charge gap.  

At  ${l_{\mu} \ll min\{l_{\Delta},l_{M_{c}}\}}$ the gap creating terms have strongly oscillating arguments and are 
wiped off upon integration, and therefore at large distances the effective theory is given by  two independent 
Gaussian fields 
\bea
 {\cal H}_{i} &=&   \sum_{i=c,s}\int dx \Big\{\frac{v_{i}}{2}\big[(\partial_{x}\phi_{i})^2
+(\partial_x \theta_{i})^2\big]\, , 
\label{Gausss}
\eea
describing the Luttinger-liquid metallic phase with gapless charge and spin excitation spectrum. In deriving Eq.\ (\ref{Gausss})
we have  taken into account that the perturbation caused by the cosine term in the spin channel is marginally irrelevant at 
{${U>0}$}.  
Thus, within the used 2-FP approximation, 
the bosonization treatment predicts the {\em commensurate-incommensurate} nature of both the BI-metal and metal-CI transitions.

In the opposite case, where ${l_{\mu} \gg max\{l_{\Delta},l_{M_{c}}\}}$, "doping" is ineffective and may be neglected.  
The corresponding effective field theory coincides with that of the standard IHM~\cite{Fab_99} 
\emph{i.e.\ } the theory of the two Gaussian fields in Eq.\ (\ref{Gausss}) coupled by the effective potential
\bea
&&
\hspace{-10mm} V_{cs}  = 
\frac{M_{s}}{2 \pi^{2} a_{0}^{2}}\cos\sqrt{8\pi}\phi_{s}
+
\frac{M_{c}}{2 \pi^{2} a_{0}^{2}}\cos\sqrt{8\pi K_{c}}\phi_{c}\nonumber \\ 
&&
-
\frac{\Delta_{r}}{\pi a_{0}}\sin\sqrt{2\pi K_{c}}\phi_{c} \cos\sqrt{2\pi}\phi_{s},
\label{H_CS_BOS_1} 
\eea
where $M_{c}$ and $M_{s}$ are considered as  phenomenological parameters characterizing charge and spin gaps. 
In the gapped regime fluctuations of the corresponding fields are suppressed and 
the properties of the system are determined by the vacuum expectation values of the fields $\phi_{s}$ and $\phi_{c}$, 
which correspond to the minimum of the potential energy  in Eq.\ (\ref{H_CS_BOS_1}). 
Below in our analysis we follow the route developed in Ref.~\cite{Fab_99}.

At weak $U$, where $l_{\Delta} < l_{\mu} \ll l_{M_{c}}$ is the shortest length scale in the theory, 
the minimum of the potential energy is reached at the following two sets of minima (defined modulo $2\pi$): 
$\langle \phi_{s} \rangle = 0$,  $\sqrt{2\pi K_{c}}\langle\phi_{c}\rangle=\pi/2$  
and 
$\langle \sqrt{2\pi}\phi_s\rangle  = \pi $, $\langle \sqrt{2\pi K_{c}}\phi_c \rangle = -\pi/2$. 
These sets characterize the BI phase. 
Indeed in this case the alternating on-site charge density operator
\bea
&{\cal Q} (x)=(-1)^{i}n_{i}\sim \sin \sqrt{2\pi K_{c}}\phi_c\cos \sqrt{2\pi}\phi_s&
\eea
acquires a finite vacuum expectation value. Moreover, the vacuum-vacuum  transitions, $\Delta \phi_{s(c)} = \pm \pi$, describe stable topological excitations carrying the charge $Q=\Delta\phi_c /\pi =\pm 1$ and spin $S^z =\Delta \phi_s / 2\pi = \pm 1/2$  and therefore coinciding with ``massive'' single-fermion excitations of the BI. 

At strong repulsion, where the large correlated (Hubbard) charge  gap $l_{M_{c}}< l_{\mu} \ll l_{\Delta}$ determines the shortest length-scale of the system, 
the situation changes and each minimum in the charge sector splits into two degenerate minima: 
$\langle\phi_s \rangle= 0$, $\langle\sqrt{2\pi K_{c}}\phi_c \rangle = \phi_0$,  $\pi -\phi_0$,
and 
$\langle\sqrt{2\pi}\phi_s \rangle = \pi$, $\langle\sqrt{2\pi K_{c}}\phi_c \rangle= -\phi_0$, $-\pi +\phi_0$,
where 
$$
\phi_0 = \arcsin (\pi \Delta_{r}/2M_{c}).
$$ 
These new sets of minima support, besides the CDW order, also the BOW order because for $\langle\sqrt{2\pi K_{c}}\phi_c \rangle \neq \pm \pi/2$ the dimerization operator
\bea
{\cal D}(x)&=&\sum_{\sigma}(-1)^{n}(c^{\dagger}_{i,\sigma} c\PHDG_{i+1,\sigma}+ \mathrm{H.c.})\nonumber\\
&\sim& \cos\sqrt{2\pi K_{c}}\phi_c (x)\cos\sqrt{2\pi}\phi_s (x)
\eea
acquires a finite expectation value in the new vacuum. 
The location of the minima in the spin sector, and hence the spin quantum numbers of the topological excitations, 
are the same as in the BI phase. 
However, the charge quantum numbers become \emph{fractional}, depending on $\phi_0$. 
The $Z_2$-degeneracy of the spontaneously dimerized state implies the existence of topological kinks carrying the spin S = 1/2 and charge  $Q = \pm 2\phi_0 /\pi$ ~\cite{Fab_99}. 

Thus  eventually, with increasing Hubbard repulsion, at ${l_{\Delta} \simeq l_{M_{c}}}$ the BOW pattern is generated in the ground state. If the transition takes place at ${l_{\Delta} \simeq l_{M_{c}}< l_{\mu} }$ \emph{i.e.\ }within the gapped phases one recovers the phase diagram of the standard IHM~\cite{Fab_99}. However, if the same transition takes place at 
${l_{\mu} < l_{\Delta} \simeq l_{M_{c}}}$ \emph{i.e.\ }in the metallic phase, although the charge excitation spectrum is gapless, in the ground state coexistence of the LRO CDW and BOW patterns will be present. 

\subsection{Large $U$ spin chain limit \label{subsec:Heisenberg limit}}

To complete our qualitative analysis, notice that the 
behavior of the spin gap substantially depends on the value of the parameter ${t^{\prime}/t}$. 
At strong repulsion ${U\gg t,t^{\prime},\Delta}$ the spin degrees of freedom are described by the Hamiltonian of frustrated Heisenberg chain 
\begin{equation}
{\cal H}_\text{Heis}= J\sum_{n} {\bf S}_{n} \cdot {\bf S}_{n+1} +
J^{\prime}\sum_{n} {\bf S}_{n} \cdot {\bf S}_{n+2}\, ,
\label{J-J'-HeisChain}
\end{equation}
where \cite{Grusha_2016} 
\begin{eqnarray}
	J &=&\frac{4t^{2}}{U}\left[1-\frac{1}{U^{2}}\left(4t^{2}-\Delta^{2}\right)\right]+\mathcal{O}(1/U^{5}),\\
	J^{\prime}&=&\hspace{-1.5mm}\frac{4t'^{2}}{U}\left[1-\frac{1}{U^{2}}\left(\frac{4t'^{4}-t^{4}}{t'^{2}}\right)\right]+\mathcal{O}(1/U^{5}).
\end{eqnarray}
Excitation spectrum of the spin chain (\ref{J-J'-HeisChain}) is gapless at ${J^{\prime}/J<1/4}$ and 
gapped at ${J^{\prime}/J> 1/4}$~\cite{Haldane_82,Okamoto_92}. 
Consequently, at large $U$  and ${t^{\prime}<0.5\,t}$ the spin excitation spectrum is gapless, 
while at ${t^{\prime}>0.5\,t}$ -- is gapped. 
Hence, at ${t^{\prime}<0.5\,t}$ with increasing $U$ after the appearance of the BOW phase the spin gap closing transition takes place ~\cite{Fab_99}, 
while at ${t^{\prime}>0.5\,t}$ the spin gap remains finite in the whole area of the CI phase even at large ${U}$.

\section{Numerical exploration \label{sec:Numerics}}

 In order to test the validity of the picture obtained in the previous Section
we investigated numerically the predicted insulator--metal--insulator transitions 
and relevant order parameters in the different phases.
To this end we have performed  DMRG \cite{White_1992} calculations on finite length $L$ chains with open boundary conditions (OBC).
The employed code relies on the ITensor software library \cite{White_2020}.

The parameter region of interest, as described in Section \ref{sec:Introduction}, 
is the full range of Hubbard repulsion $U>0$ in the close proximity of the insulator-metal 
(Lifshitz) transition of the $t-t'$ ionic chain. 
This is achieved  with $t' \lesssim t'_c$, 
where we expect to find a band insulator phase (induced by $\Delta$ at low $U$), 
a metallic phase at intermediate $U$ (induced by second neighbor hopping amplitude $t'>0.5t$),
and a correlated insulator phase for large $U$.
We found it convenient to set the energy scale as $t=1$, to choose $\Delta=0.8$  and $t'=0.55$ (being $t'_c \approx 0.638$), 
exploring the effects of Hubbard repulsion $U$ from the non-interacting regime ($U=0$) 
up to large enough values to reach a Mott-like insulator, estimated as $U \sim 4.0$ . 

As the Hamiltonian ${\cal H}$ in Eq.\ (\ref{eq:t1t2_IH_model}) commutes with the total number operator ${\cal N}={\cal N}_\uparrow+{\cal N}_\downarrow$
and the total magnetization operator 
${\cal S}^z=\left({\cal N}_\uparrow-{\cal N}_\downarrow\right)/2\, ,$
one can compute the lower eigenvalue states of ${\cal H}$ within subspaces 
with given quantum numbers $N$ for the number of electrons and $S^z$ for the total spin projection. 
We then denote  by $E_{0}(N,S^{z})$ the lowest eigenvalue and by $E_{1}(N,S^{z})$ the first excited eigenvalue in the given subspace.

Specifically, we have focused on the following states (notice that, because of spin symmetry, reversing the sign of $S^z$ does not change the eigenvalues):

\begin{itemize}
	
	\item  $N=L$, $S^{z}=0$, the ground state with lowest eigenvalue $E_{0}(L,0)$ and the ``internal excitation'' with first excited eigenvalue $E_{1}(L,0)$
	
	\item $N=L$, $S^{z}= 1$, the ``spin flip'' state with lowest eigenvalue $E_{0}(L, 1)$ 
	
	\item $N=L+1$, $S^{z}=+1/2$, a ``one particle'' state  with lowest eigenvalue $E_{0}(L+1,1/2)$ 
	
	\item $N=L-1$, $S^{z}=+1/2$, a ``one hole'' state  with lowest eigenvalue $E_{0}(L-1,1/2)$
	
	\item $N=L+2$, $S^{z}=0$, the ``two particle'' state  with lowest eigenvalue $E_{0}(L+2,0)$
	
	\item $N=L-2$, $S^{z}=0$, the ``two hole'' state  with lowest eigenvalue $E_{0}(L-2,0)$
	
\end{itemize}

These states where computed using maximal bond dimensions up to 800, the truncation error being lower than $10^{-8}$.   
However, when the energy difference between $E_{1}(N,S^{z})$ and $E_{0}(N,S^{z})$ is too small 
DMRG convergence towards the ground state becomes difficult. 
Such difficulties indeed arose in the presumably metallic region, expected to be gapless in the thermodynamic limit, 
as we increased the chain length. Within our resources, for some values of $U$, 
we could not ensure convergence for chains beyond a hundred sites. 
Moreover, the size scaling behavior with inverse length $1/L$ might change at some critical length \cite{Manmana_04} 
making it uncertain any extrapolation technique from moderate lengths into the thermodynamic limit. 
We do not attempt in the present work to provide precise extrapolations.
We limit ourselves to show confident finite size data, 
adding suggested thermodynamic extrapolations only when the scaling tendency with $1/L$ seems stable. 
We find that the suggested results support the validity of our analytical predictions, 
as described schematically in Fig.\ \ref{fig:phase_diagram}.

The square of the total spin operator 
$\mathbf{\cal S}=\sum_{i}\left(  c^{\dagger}_{i,\sigma} \dfrac{\mathbf{\sigma}_{\sigma \sigma'}}{2} c\PHDG_{i,\sigma'} \right)$ 
also commutes with the Hamiltonian, then the total spin $S$ is a good quantum number. 
However it is not additive and can not be fixed along DMRG sweeps.
We have computed, for each state obtained, the expectation value 
$\langle \mathbf{\cal S}^2  \rangle$ 
to check coincidence with $S(S+1)$ for a given integer or half-integer $S$. 

In this sense we have found that, for any considered repulsion $U$ and length $L$, 
the half-filled, non magnetized ground state is a singlet state with $S=0$.
The internal excitation and the spin flip states form a triplet with $S=1$.  
Consistently with spin symmetry they are degenerate, $E_{1}(L,0)=E_{0}(L,\pm 1)$. 
This is the lowest excitation of the ground state. 
We have found no signal of other exciton state lying below the spin triplet,
in contrast with the situation observed in the nearest neighbors IHM \cite{Manmana_04}. 

For the ground state we have also computed the local charge and spin densities, and spin correlations along the chains, 
with the aim of discussing order parameters in the different phases.

We describe below the results of different measures we have performed, setting $t=1$, $t'=0.55$ and $\Delta=0.8$, 
on chains of several lengths up to $128$ sites.

\subsection{Energy gaps\label{subsec:Energy gaps}}

One can define different gaps with respect to the half-filled  ground
state, corresponding to the different possible excitations. We consider the following:

\begin{itemize}
	\item the internal gap  $\Delta_{int}$ in the subspace with $N=L$ and $S^z=0$,
	\begin{equation}
	\Delta_{int}=E_{1}(L,0)-E_{0}(L,0) \label{gap-int}
	\end{equation}
	\item the spin gap $\Delta_{s}$ corresponding to spin flipped states $S^z=\pm 1$ with $N=L$,
	\begin{equation}
	\Delta_{s}=\frac{E_{0}(L,1)+E_{0}(L,-1)-2E_{0}(L,0)}{2} \label{gap-spin}
\end{equation}
	\item the one-particle gap $ \Delta_1$ corresponding to the addition/subtraction of one electron,
	\begin{equation}
	\Delta_1=E_{0}(L+1,1/2)+E_{0}(L-1,1/2)-2E_{0}(L,0) \label{gap-1}
\end{equation}
	\item the two-particle gap $\Delta_{2}$ corresponding to the addition/subtraction of charge while keeping the magnetization $S^z=0$,
	\begin{equation}
	\Delta_{2}=\frac{E_{0}(L+2,0)+E_{0}(L-2,0)-2E_{0}(L,0)}{2} \label{gap-2}
	\end{equation}
\end{itemize}
Notice that a chemical potential should be added to ensure that the half-filling $N=L$ sector contains the ground state of the system.
%
%
However chemical potential contributions cancel out in these
gap constructions, then gaps can be computed directly from the eigenvalues
of the Hamiltonian in Eq.\ (\ref{eq:t1t2_IH_model}.)

From the degeneracy of the spin triplet one can see that $\Delta_{int}=\Delta_{s}$. Moreover,
the present definition of the spin gap coincides with 
the difference between the triplet and singlet energies at half-filling 
($E(N=L, S=1) - E(N=L,S=0)$) used elsewhere. 
From the same relation, as there is no exciton state below the spin gap, 
we assume that $\Delta_{2}$ is a meaningful measure of the charge gap. 
We denote $\Delta_{2}$ as $\Delta_{c}$ in the following.

\begin{figure}[h] 
	\begin{centering}
		\includegraphics[scale=0.6]{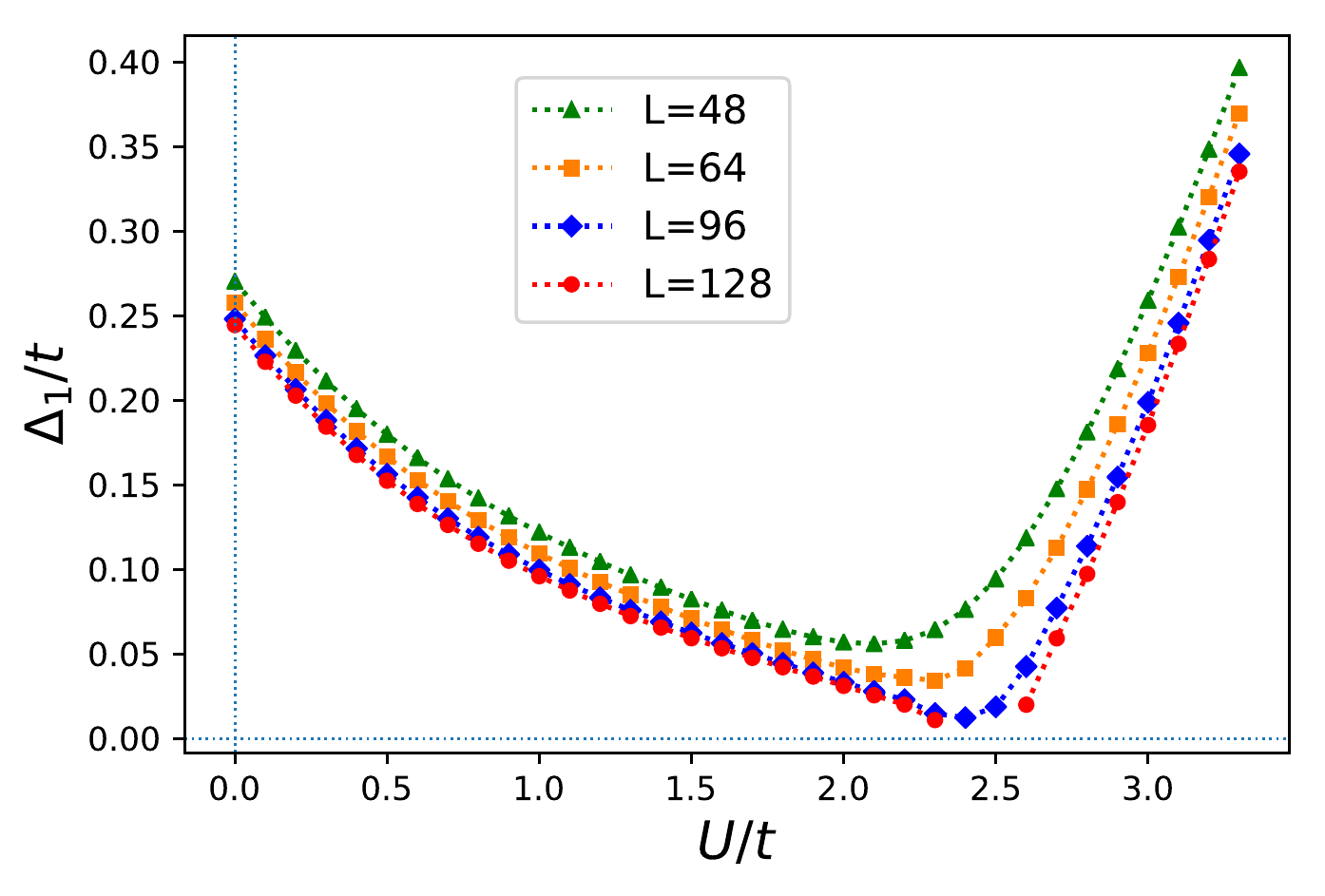}
		\par\end{centering}
	\caption{
		One-particle gaps $\Delta_{1}$, for $t'=0.55\,t$ and $\Delta=0.8\,t$.
		Data from finite chains of different lengths $L=48,64,96, 128$ is shown 
		(some points for $L=128$ are not included).
		One can distinguish the band insulator phase for low $U$,  signals of a gapless region for intermediate $U$, and  a re-entrance to a large $U$ insulator phase.
	}
	\label{fig: 1p energy gap}
\end{figure}
%

We first show in Fig.\ \ref{fig: 1p energy gap} the one-particle gap $\Delta_{1}$,
which involves the change of both charge and spin quantum numbers.
The key feature of this plot is the  apparent presence of a gapless region for intermediate $U$. 
Notice that some points for $L=128$ with convergence difficulty are not included;
in these cases the gap seems to be so small that our procedures have not been able 
to separate the ground state from the first excited level. 

\begin{figure}[h] 
	\begin{centering}
		\includegraphics[scale=0.6]{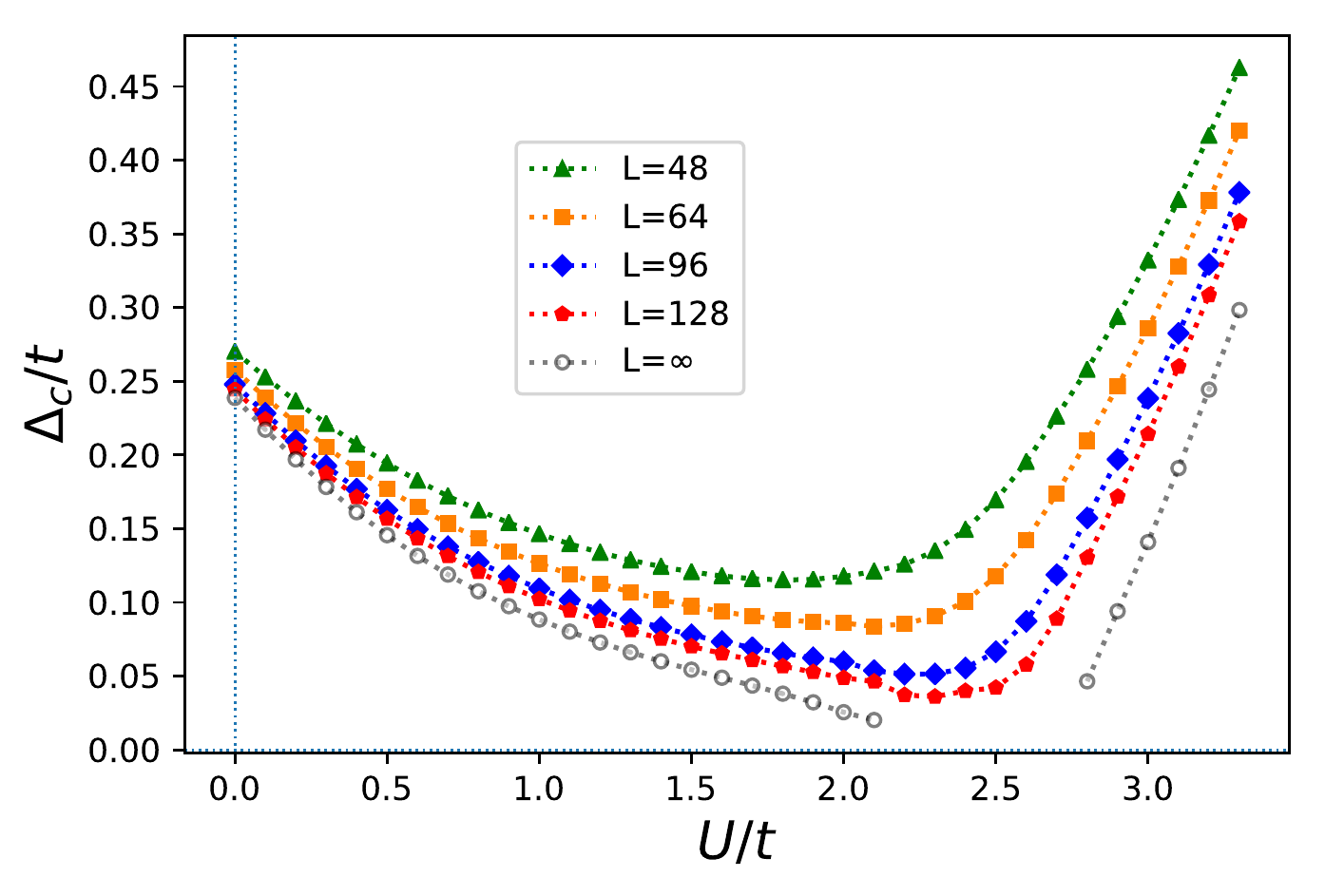}
		\par\end{centering}
	\caption{
		Two-particle charge gap $\Delta_{c}$, for $t'=0.55\,t$ and $\Delta=0.8\,t$.
		Data from finite chains of different lengths $L=48,\, 64,\, 96,\, 128$ is shown, 
		together with a proposed extrapolation where appropriate (hollow circles).
		The band insulator phase for low $U$ and the correlated insulator phase for large $U$ can be distinguished. 
		At intermediate $U$ the lengths computed do not provide a definite scaling tendency; 
		we argue in the Fig.\ \ref{fig: 2p energy gap scaling x3} that our data is consistent with a gapless
		thermodynamic limit.			
	}
	\label{fig: 2p energy gap}
\end{figure}
%

\begin{figure}[h] 
	\begin{centering}
		\includegraphics[scale=0.6]{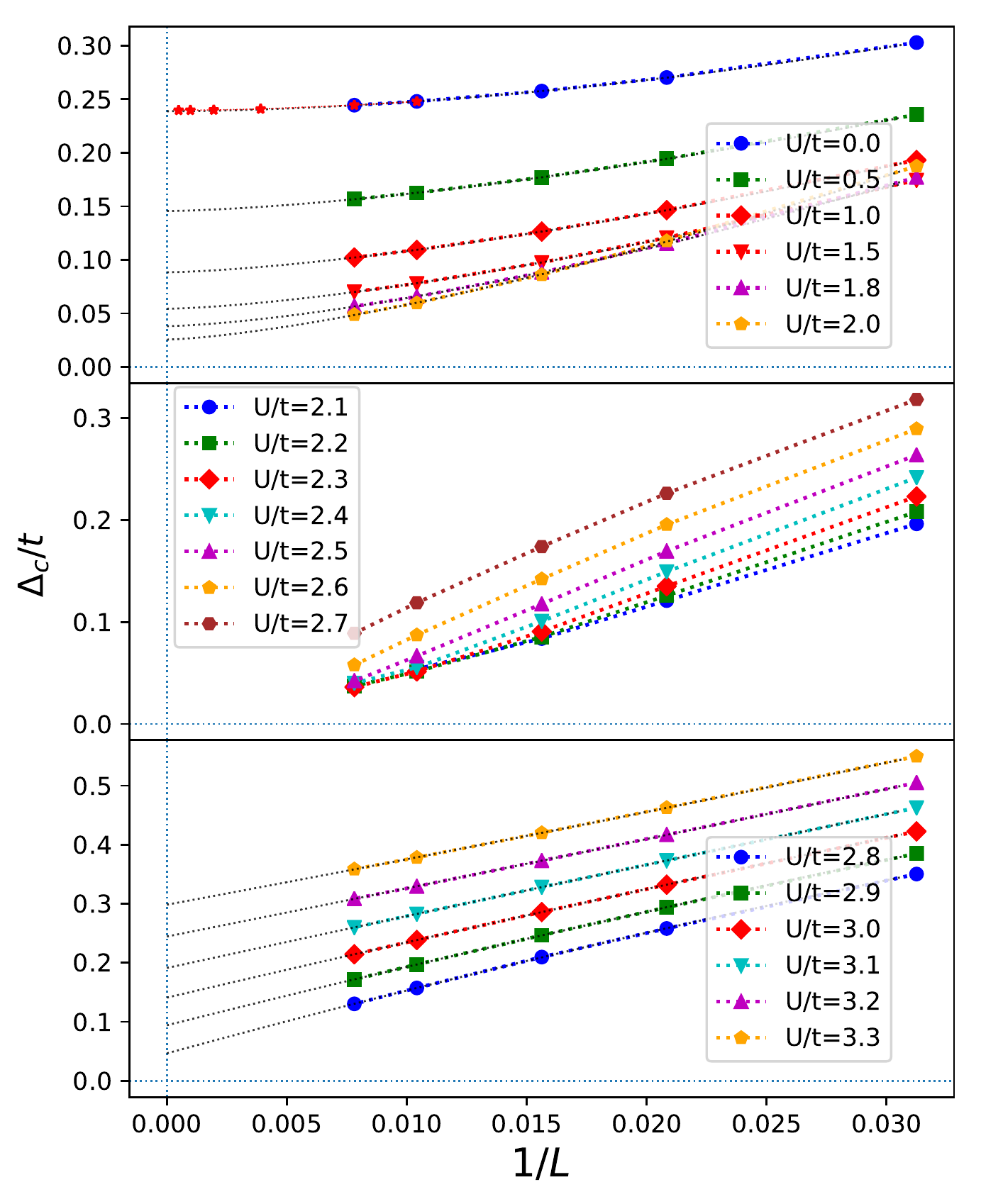}
		\par\end{centering}
	\caption{
		Finite size scaling of the charge gap for different values of the Hubbard repulsion, with $L$ ranging from 32 to 128 sites. An extrapolation is shown as a guide to the eyes when appropriate.
		Top panel: for low $U$ the charge gap scaling can be fitted with a power law $\Delta_c(\infty)/t+L^{-\nu}$, and clearly extrapolates towards a non zero band insulator gap (for $U=0$ we added large size free-electron results, in red stars). 
		 Middle panel: in the intermediate region the scaling concavity changes from positive to negative. 
		A naive extrapolation from our finite size data is misleading,
		meaning that there should be a change in the scaling tendency at larger lengths. 
		Though refined computations are needed, a graphical inspection strongly suggests that 
		present results are consistent with a gapless thermodynamic limit.
		Bottom panel: for larger $U$ the negative scaling concavity smoothly gives place to a polynomial behavior. 
		For $U\geq 2.8\,t$ a quadratic extrapolation is again clearly non zero, corresponding to the correlated (Mott-like) insulator phase.
	}
	\label{fig: 2p energy gap scaling x3}
\end{figure}
%

\begin{figure}[h] 
	\begin{centering}
		\includegraphics[scale=0.6]{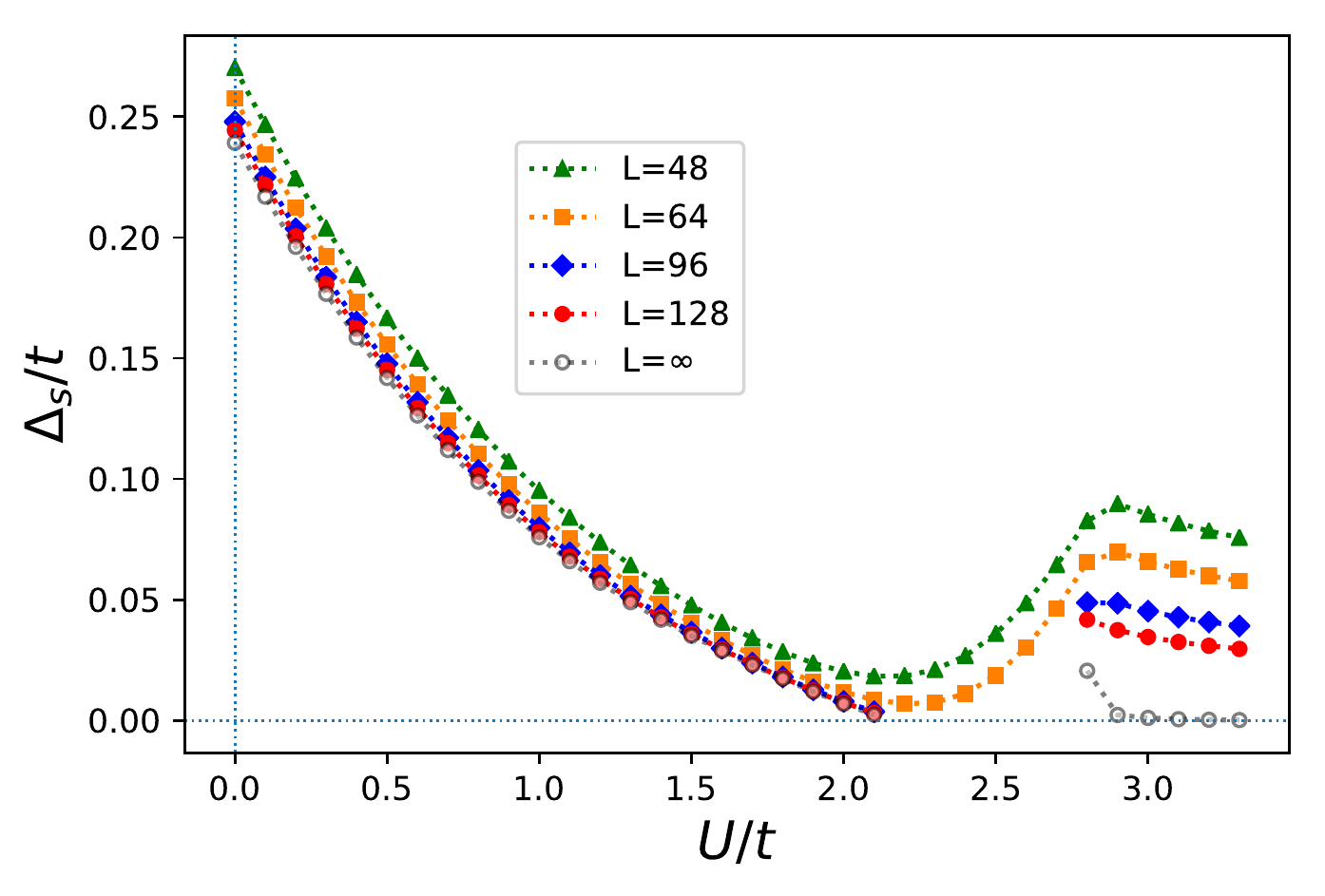}
		\par\end{centering}
	\caption{
		Spin gap $\Delta_{s}$, for $t'=0.55\, t$ and $\Delta=0.8\, t$.
		Data from finite chains of different lengths $L=48,\, 64,\, 96,\, 128$ is shown 
		only where DMRG convergence is reached. 
		The band insulator phase for low $U$ with similar spin and charge gaps can be distinguished. 
		The correlated insulator phase for large $U$ shows a rise of the spin gap followed by a slow decay.
		An extrapolation is shown when appropriate (hollow circles, see details in Fig.\ \ref{fig: 2p spin gap scaling x3}).
	}
	\label{fig: 2p spin gap}
\end{figure}
%

In order to analyze separately charge and spin  degrees of freedom 
we show in Fig.\ \ref{fig: 2p energy gap} the two-particle charge gap $\Delta_{c}$ ($\Delta_2$). 
As expected for finite systems \cite{Manmana_04}, we observed that $\Delta_{c}>\Delta_{1}$.
The existence of a gapless region at intermediate $U$, in the  thermodynamic limit, is not evident 
from the largest length studied and requires a detailed size scaling analysis. 
In Fig.\ \ref{fig: 2p energy gap scaling x3} we show that the $1/L$ scaling behavior is very different at low, mid or large $U$.
A power law $L^{-\nu}$ in the BI phase, and a quadratic polynomial in the CI phase, 
fit well the finite size data providing the suggested extrapolation in Fig.\ \ref{fig: 2p energy gap} (in gray). 
However, in the region $2.2 \lesssim U \lesssim 2.7$ it is apparent that larger sizes are needed to define $1/L$ scaling.
Though we do not propose an extrapolation, a graphical inspection suggests the presence of the unusual gapless phase in this region.


\begin{figure}[h] 
	\begin{centering}
		\includegraphics[scale=0.6]{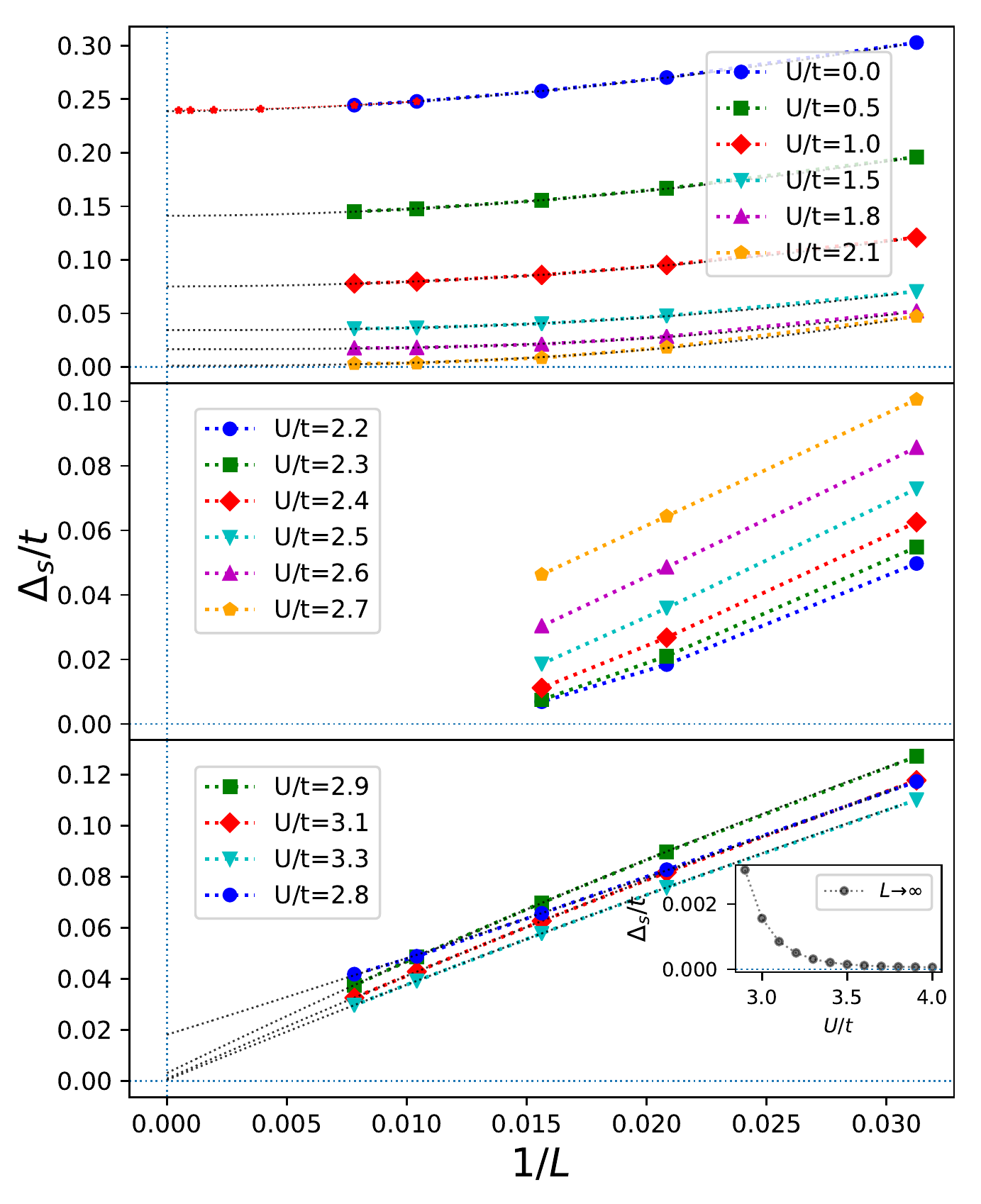}
		\par\end{centering}
	\caption{
		Finite size scaling of the spin gap for different values of the Hubbard repulsion, with $L$ ranging from 32 to 128 sites. 
		Top panel: for low $U$ a power law scaling of the spin gap suggests  non zero extrapolations, with values similar to the charge gap (for $U=0$ large size free-electron results are also shown, in red stars). 
		The extrapolated spin gap decreases smoothly with $U$, while the scaling maintains the slope and concavity. We estimate that it is non zero up to $U\approx 2.1\,t$.  
		Middle panel: in the intermediate region the scaling looks almost linear, 
		but a naive extrapolation would lead to meaningless results; 
		this means that for larger lengths there should be a cross-over in the scaling tendency. 
		Though we have not reached DMRG convergence for larger systems in this region, 
		the behavior might be compatible with a gapless thermodynamic limit up to $U\approx 2.7\,t$. 
		Bottom panel: a singular behavior is observed at $U=2.8\,t$, where the spin gap re-opens and gets a peak value. For higher $U\geq 2.9\,t$ the scaling gets a slight negative concavity and a quadratic extrapolation decreases smoothly towards zero. Inset: a quadratic extrapolation in this region suggests non-vanishing, decaying, spin gaps.
		}
	\label{fig: 2p spin gap scaling x3}
\end{figure}
%

Next we show in Fig.\ \ref{fig: 2p spin gap} the spin gap $\Delta_{s}$, 
coincident with the internal excitation gap in the half-filled, non-magnetized subspace of states.  
Being the lowest excitation of the ground state, we have not reached good DMRG convergence 
in the $2.2 \leq U \leq 2.7$ region  
where the internal excitation could not be separated from the ground state.  
From the available data we show in Fig.\ \ref{fig: 2p spin gap scaling x3} the scaling tendency. 
One finds a finite spin gap in the band insulator region, 
a possibly spin gapless phase in the intermediate region and a re-opening of the spin gap in the correlated insulator region.
In this last region we observed a regular scaling behavior that leads to a sensible mathematical extrapolation: 
a quadratic fit provides a small but non-vanishing, decaying, spin gap in the thermodynamic limit (shown in the inset). 
This is consistent with the spin dimerized phase predicted in Section \ref{subsec:Heisenberg limit}.
The suggested extrapolation is plotted in Fig.\ \ref{fig: 2p spin gap} (in gray).
Further investigation, exceeding our numerical resources, is needed in the intermediate region.
From the shown data one can infer for low $U$ a band insulator type region
(BI, non correlated))  with (almost) $\Delta_{c}=\Delta_{s}$.
The gaps decay as the repulsion $U$ penalizes double occupation of
low-potential (odd) sites and promotes n.n.n. hopping $t'$ between high-potential (even) sites. 
The charge gap $\Delta_{c}$ and the spin gap $\Delta_{s}$ presumably close 
at $U_{c,1} \approx 2.2$ (we can not resolve whether they would close at the same point),
giving rise to the repulsion driven metallic phase. 
When larger repulsion $U$ gets strong enough to also penalize double occupation of
high-potential sites the charge gap re-opens 
and starts to grow  with $U$. This occurs at $U_{c,2} \approx 2.7$. 
It is expected that the charge gap increases linearly in this region,
from the fact that our computations are done with a fixed number of particles
instead of fixing the chemical potential 
(see  \cite{Vekua_2009} and Appendix \ref{app: B} for a discussion). 
Interestingly, the spin gap also re-opens close to  $U_{c,2}$, 
and grows to a maximum in a narrow range of $U$, as if bound to the charge gap. 
This unusual behavior seems not to be captured by the 2-FP bosonization approach in Section \ref{sec:bosonization}.
Beyond a peak value at $U \approx 2.80$ the spin gap  starts to decay while the charge gap keeps growing, 
signaling a strongly correlated insulator phase.


\begin{figure} 
	\begin{centering}
		\includegraphics[scale=0.55]{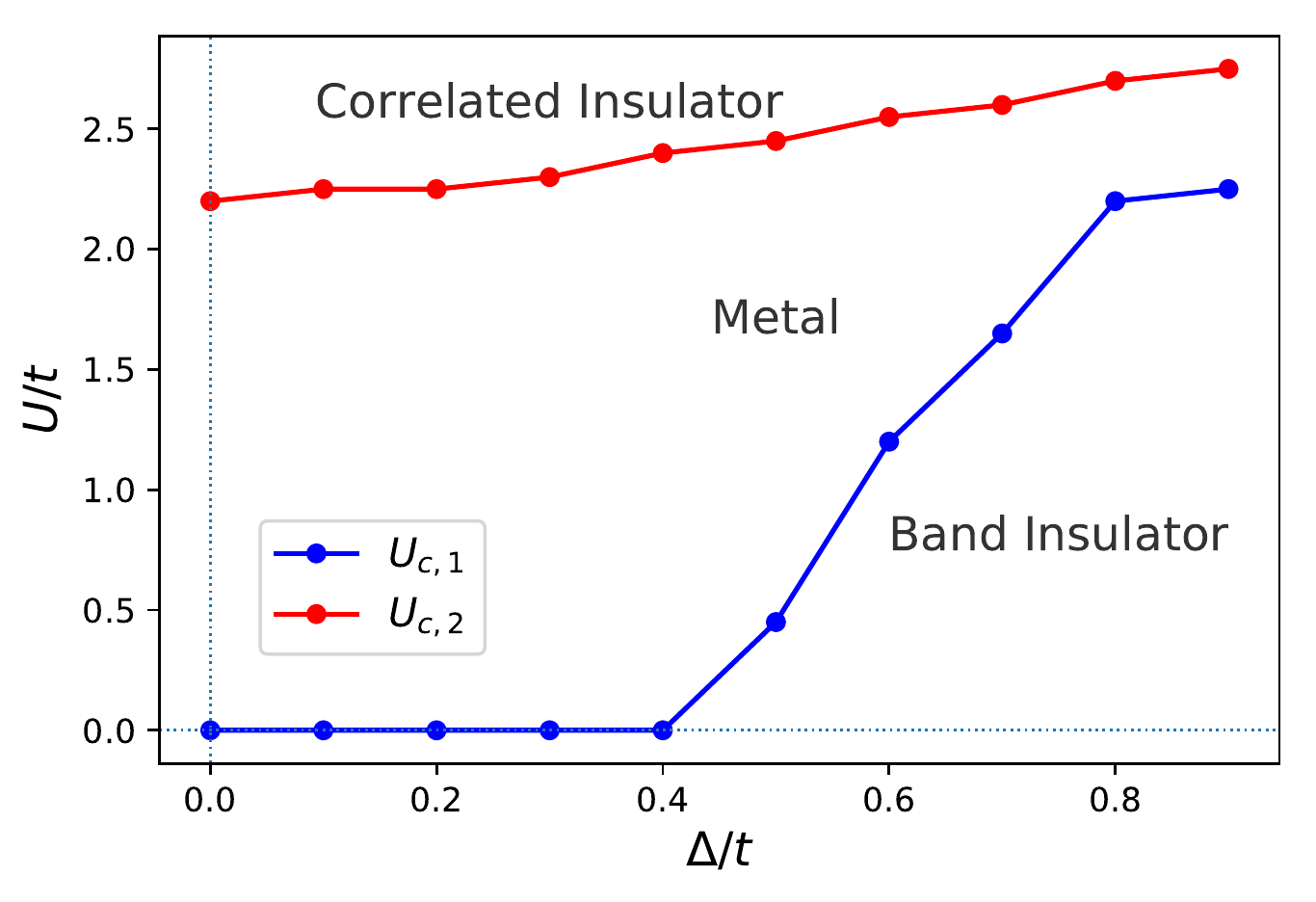}
		\par\end{centering}
	\caption{
		Estimated transition points for different ionicities $\Delta$ 
		and $t'=0.55\, t$. $U_{c,1}$ corresponds to the band insulator -- metal transition where the charge gap vanishes, 
		$U_{c,2}$ to the metal -- correlated insulator transition  where the charge gap re-opens.   
	}
	\label{fig:U-Delta}
\end{figure}
In order to investigate the role of the ionicity $\Delta$ in the gap formation 
we have additionally explored  the range $0 \leq \Delta \leq 0.9$, keeping $t=1$ and $t'=0.55$ close to the Lifshitz point.
Without reaching further numerical precision, 
we have observed that when the ionic potential amplitude $\Delta$ is lower the argued metallic region starts at lower $U_{c,1}$ 
and is eventually present since the free point $U=0$  when  $\Delta$ is low enough. 
The value of $U_{c,2}$ where the charge gap re-opens is less sensitive to the ionicity.
The spin gap peak close to  $U_{c,2}$ was observed for any  $\Delta$.
An estimation of the transition points according to the ionicity parameter is shown in Fig.\ \ref{fig:U-Delta}.

\subsection{Order parameters\label{subsec:Order parameters}}

For the computed ground states we have  evaluated the local expectation values 
$\rho_{i,\sigma}=\langle n_{i,\sigma}\rangle$ for each site 
and $q_{i,\sigma}=\langle c_{i,\sigma}^{\dagger}c_{i+1,\sigma}^{\phantom{\dagger}}+H.c.\rangle$
for each bond, as well as spin-spin correlations $\langle S^z_iS^z_j \rangle$ 
with the aim of revealing the existence of magnetic order. 
The following local densities are then considered:
\begin{itemize}

	\item local charge density $\rho_i=\rho_{i,\uparrow}+\rho_{i,\downarrow}$

	\item bond charge density $q_i=q_{i,\uparrow}+q_{i,\downarrow}$

\end{itemize}
The local spin density $\sigma_i=\frac{1}{2}\left(\rho_{i,\uparrow}-\rho_{i,\downarrow}\right)$ 
and the bond spin density $q_{i,\uparrow}-q_{i,\downarrow}$ do vanish, as expected from the $SU(2)$ symmetry of the model
and the zero magnetization condition.

\subsubsection{Charge density wave\label{subsubsec:CDW}}

\begin{figure} 
\begin{centering}
 \includegraphics[scale=0.5]{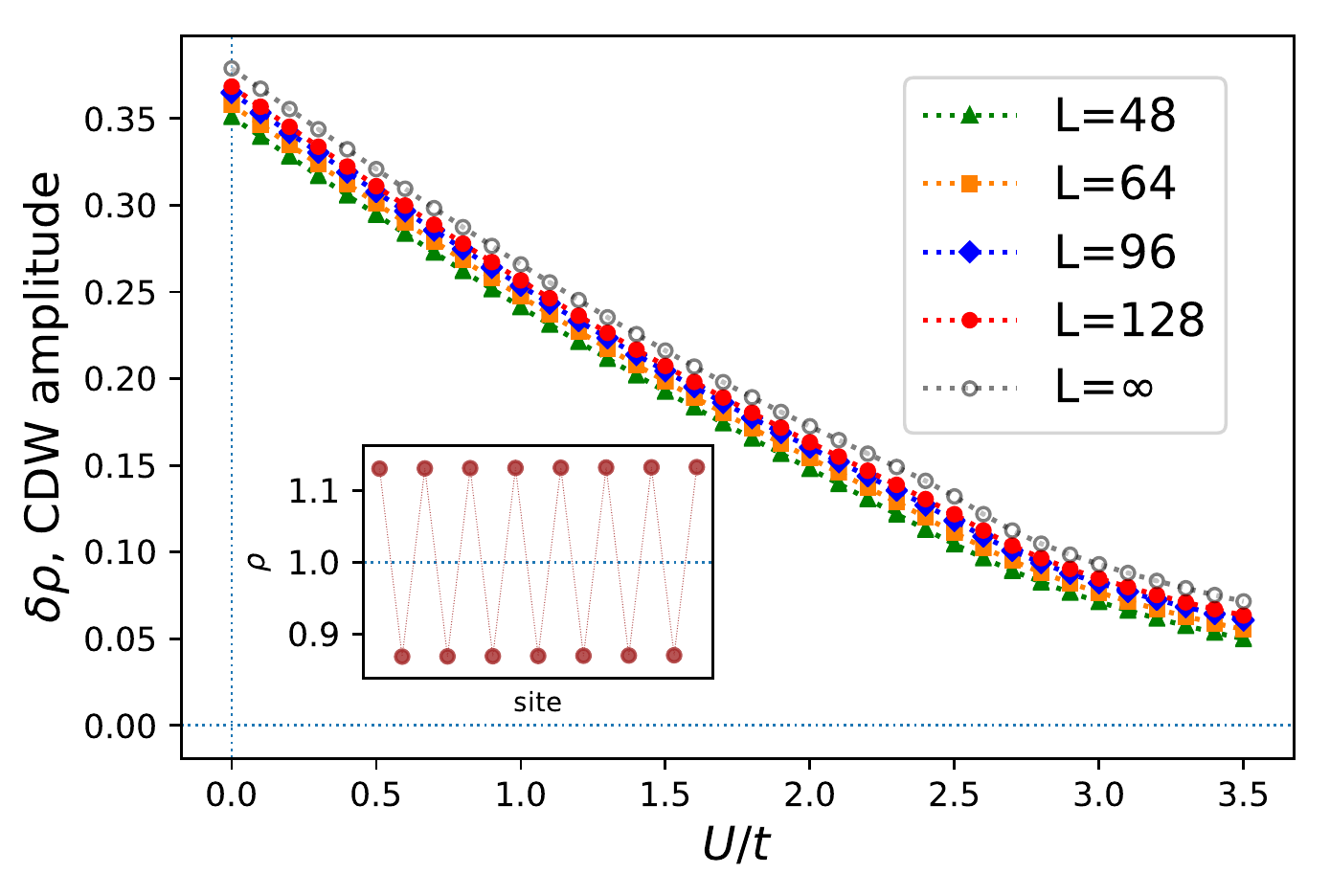}
\par\end{centering}
\caption{
	The amplitude of local charge density alternation $\delta \rho$ decreases smoothly with $U$. 
	Data is averaged along chains of length $L=48,64,96,128$ and extrapolated linearly in $1/L$.
	Notice that the slope is slightly different 
	in the metallic region. 
	Inset: detail of the CDW in a portion of a chain sample ($L=96$ sites) 
	for $U=2.5\, t$ in the metallic phase; 
	the same alternating occupation pattern is observed for all $U$.
}
\label{fig: CDW}
\end{figure}

Our results for the induced CDW order (ionicity) are summarized in Fig.\ \ref{fig: CDW}.
Local charge density $\rho_i$  is found to be alternating around the half-filling average $\bar{\rho_i}=1$, 
following the pattern induced by ionic potential. 
According to Eq.\ (\ref{eq:t1t2_IH_model}) even sites have higher local potential so they are less occupied by electrons. 
We show in the inset the charge density in the central portion of a chain sample 
($U=2.5$, $L=96$, gapless region) to illustrate the CDW order. 
A similar alternating pattern is observed in the band insulator and correlated insulator phases;
boundary effects disappear in a few sites and the occupation alternation gets homogeneous in the bulk. 

The CDW amplitude for chains of length $L$ was then computed as
\begin{equation}
	\delta\rho =  \frac{1}{L} 
	\sum_{i=1}^{L} (-1)^{i+1} \rho_{i}	
	\label{cdw}
\end{equation}
comparing the occupation of odd and even sites along the chains.
According with the short range of boundary effects, we found that the finite size scaling is linear in $1/L$. 
These results provide full support for the mean field approach developed in Section \ref{sec:MeanField}
and are
in concordance with the mean field parameter $\delta\rho_0$ defined in Eq.\ (\ref{eq:MF_n}).
We show in the main panel of Fig.\  \ref{fig: CDW} the finite size values of $\delta\rho$ 
and the corresponding extrapolation.
It is clear that
that $\delta\rho$ decreases with $U$, 
as the Hubbard repulsion penalizes local occupation larger than one 
({\em cf.} the mean field $\delta\rho_0(U)$ in Fig.\ \ref{fig:mean_field-solution}).
%

\bigskip{}

\subsubsection{Bond order wave \label{subsubsec:BOW}}

In the thermodynamic limit the Hamiltonian in Eq.\ (\ref{eq:t1t2_IH_model}) 
is symmetric under reflection with respect to a site. 
This implies that all bonds are equivalent, and one expects that the bond charge density $q_i$ should be homogeneous. 
However, a spontaneous parity symmetry breaking is known to occur in 
the ($t'=0$) IHM at intermediate repulsion $U$ \cite{Fab_99,Manmana_04}, 
manifest as a BOW phase with a two-fold degenerate, dimerized ground state 
characterized by alternating bond charge density $q_i$. 
We address in this Section the appearance of such a BOW phase in the $t-t'$ ionic Hubbard model.

\begin{figure}[h] 
	\begin{centering}
		\includegraphics[scale=0.5]{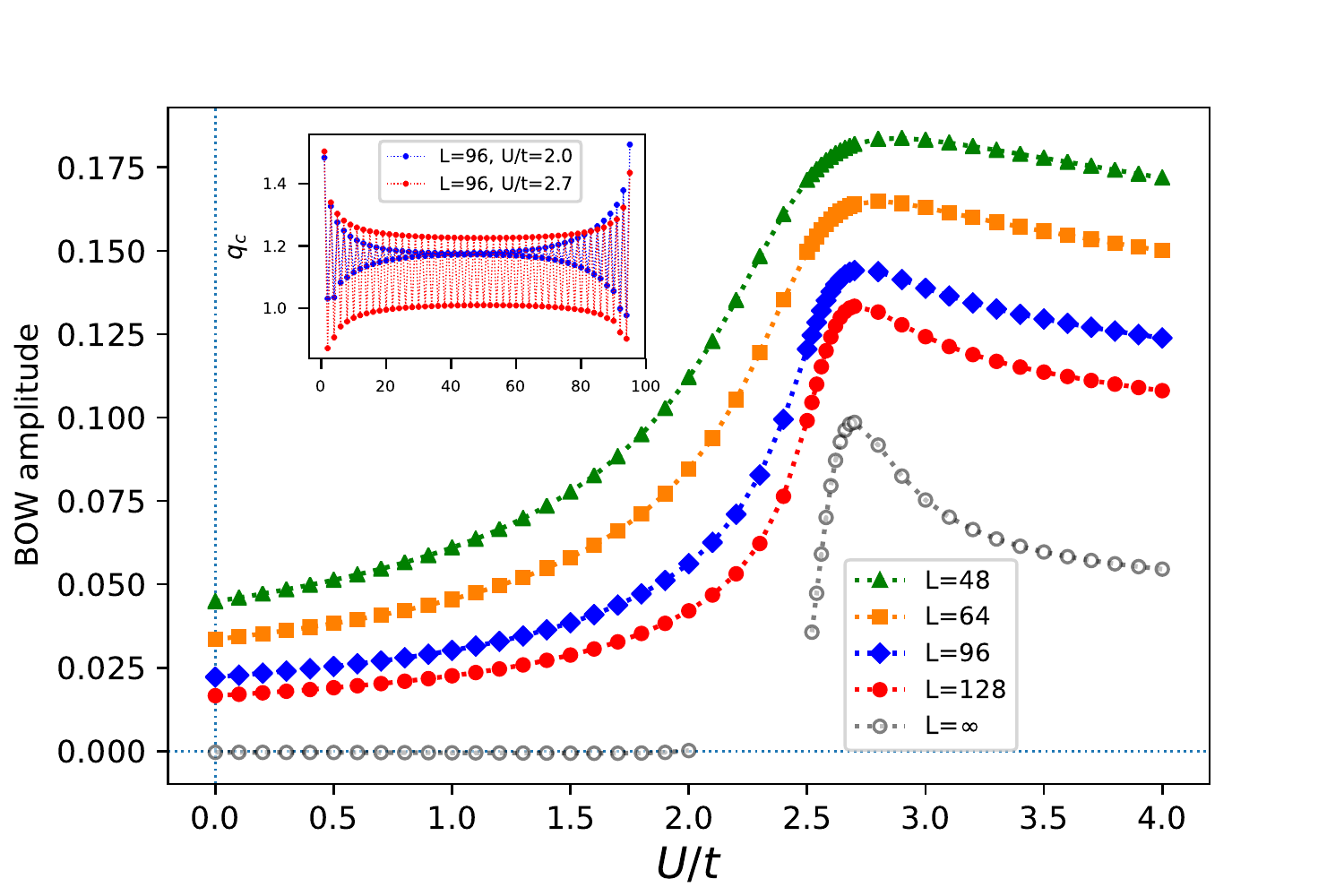}
		\par\end{centering}
	\caption{Average BOW amplitude for finite length chains with $L$ ranging from 48 to 128 sites. 
		Extrapolation to the thermodynamic limit is only suggested (hollow circles) where the scaling tendency 
		is well defined (see Fig.\ \ref{fig: BOW-scaling}). 
		No bond order is present in the BI phase
		but a  BOW amplitude appears and increases rapidly within the metallic phase ($U\geq 2.52\, t$ is shown),
		then decreases slowly in the CI phase. 
		Inset: samples of the charge bond density in a chain of length $L=96$ sites with OBC. 
		The density oscillates and the difference between odd and even bonds is always enhanced at the end bonds; 
		for the shown $U=2.0\,t$ the amplitude decays to zero towards the chain center
		but for $U=2.7\,t $  it decays to a finite steady value that signals the bulk BOW order in the $L\to\infty$ limit.
	}
	\label{fig: BOW}
\end{figure}

\begin{figure}[h]
	\begin{centering}
		\includegraphics[scale=0.5]{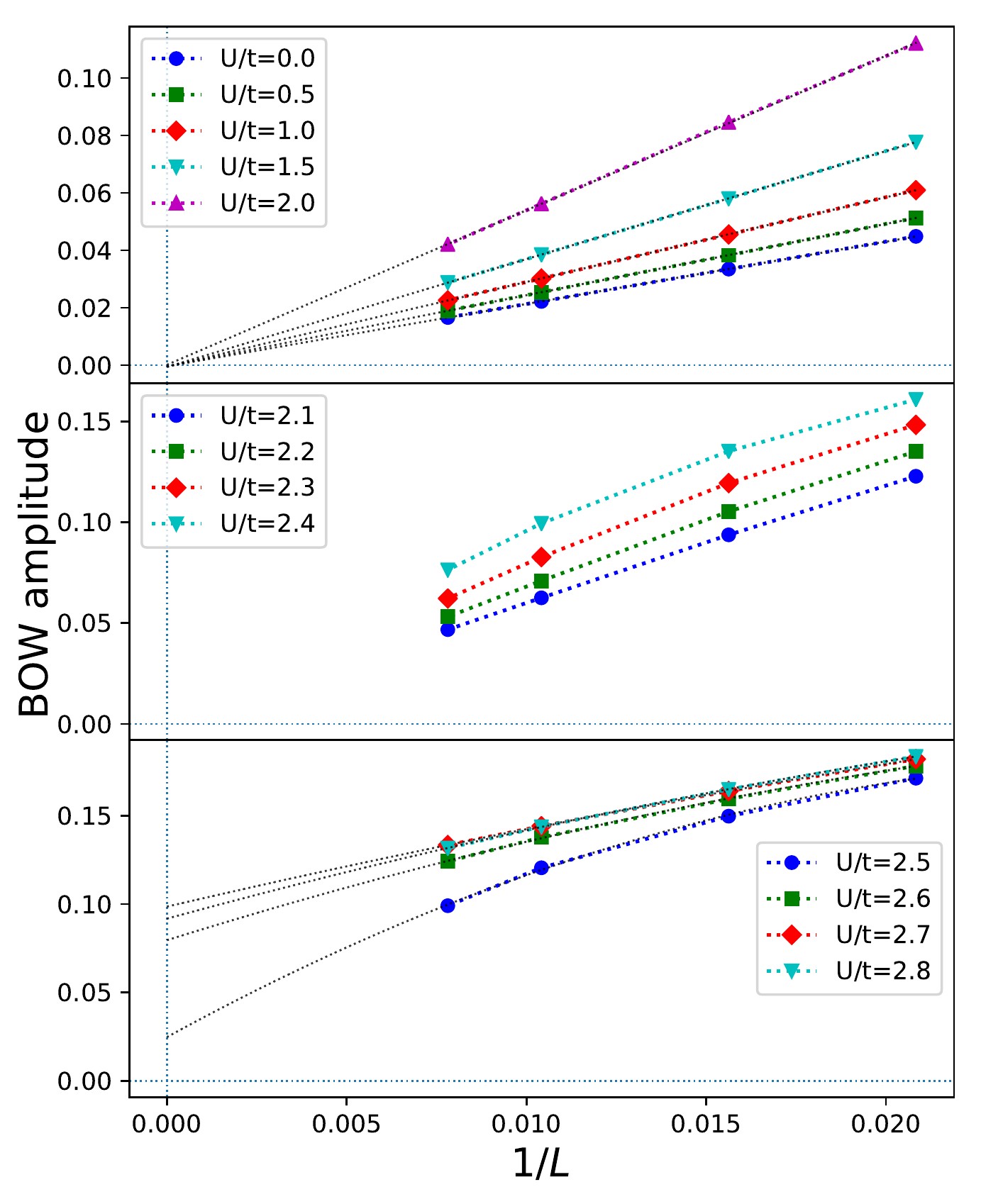}
		\par\end{centering}
	\caption{BOW amplitude scaling. 
		We show the finite size BOW amplitudes in different regions of the Hubbard repulsion, 
		and their suggested extrapolations when trustable. 
		Top panel: the average along the chains includes important boundary effects, that in the BI phase   extrapolate linearly to zero. 
		There is no bond order in this phase.
		Middle panel: in the metallic phase, up to $U=2.5\, t$ the  scaling behavior is not well defined from the computed lengths. No extrapolation is done.
		Bottom panel:  starting at  $U=2.5\, t$, within the metallic phase, 
		a quadratic extrapolation leads to non zero BOW amplitude. 
		Still, the scaling behavior at $U=2.5\, t$ might change for larger lengths.
		A maximum is reached at $U\approx 2.7\, t$, presumably coinciding with $U_{c,2}$ at the onset of the charge gap.
	}
	\label{fig: BOW-scaling}
\end{figure}

The use of OBC in the ionic chain with even number of sites $L$
explicitly breaks the reflection symmetry, 
as the edge sites have different ionic potential $\pm \Delta/2$. 
This induces an alternation of $q_i$, as shown in  
sample plots in the inset of Fig.\ \ref{fig: BOW}.
One then has to distinguish the true BOW order in the bulk from the oscillating boundary effects. 
To this end we have evaluated the average oscillation amplitudes of $q_i$ in 
the ground state of finite length chains as
\begin{equation}
	BOW =  \frac{1}{L-1} 
	\sum_{i=1}^{L-1} (-1)^{i} q_i,
	\label{bow}
\end{equation}
and then studied their scaling behavior with $1/L$. 
In Fig.\ \ref{fig: BOW} we show the finite size BOW amplitudes for a wide range of $U$
and suggest the extrapolated values where we find them trustable, from
the analysis of the scaling behaviors provided in Fig.\ \ref{fig: BOW-scaling}. 
In the BI phase the behavior is linear, leading to the absence of BOW order. 
Our present data is not enough to resolve the scaling behavior in the intermediate region, 
where the curvature can  not be clearly fitted. 
Starting within the gapless region, and extending into the correlated insulator phase, 
a quadratic extrapolation clearly indicates BOW order.
 From this analysis we suggest that the BOW order starts at some $U^\star_c$ located between $2.5$ and $2.6$, 
	and has a peak value where the charge gap re-opens. 
	Such a profound manifestation within the charge and spin gapless phase of the corresponding quantum phase transition at $U_c^\ast$  
	makes this metallic state highly unusual.
This main result is indicated in the schematic phase diagram in Fig.\ \ref{fig:phase_diagram}.

The BOW order remains present in the CI phase, with an amplitude that decreases with $U$.
As discussed in Section \ref{sec:bosonization} we do expect this  remnant BOW order,
as in the large $U$ limit the Hamiltonian in Eq.\ (\ref{eq:t1t2_IH_model}) can be mapped onto 	
a $J-J'$ spin $S=1/2$ Heisenberg model with large enough $J'>J/4$ as to be in the dimerized regime.

\subsubsection{Spin dimerization and antiferromagnetic order \label{subsubsec:SDW}}

As the expectation value of spin components vanish at every site,
the magnetic order is investigated by means of the correlation functions $\langle S^z_iS^z_j \rangle$ 
with 
\begin{equation}
S^z_i=(c^{\dagger}_{i,\uparrow} c\PHDG_{i,\uparrow}-c^{\dagger}_{i,\downarrow} c\PHDG_{i,\downarrow})/2. 
\label{Sz}
\end{equation}

On general grounds, the Hubbard repulsion $U>0$ induces antiferromagnetic correlations.
The nearest neighbor spin correlations $\langle S^z_iS^z_{i+1} \rangle$ 
might be expected to be homogeneous in the thermodynamic limit because of the site reflection symmetry;
however, spontaneous spin dimerization is known to occur in antiferromagnetic $J-J'$ spin chains \cite{Haldane_82,Okamoto_92}.
The n.n.n.\ hopping terms $t'$ in the present model introduce antiferromagnetic n.n.n.\ spin couplings
that could induce such effect.

\begin{figure}  
	\begin{centering}
		\includegraphics[scale=0.55]{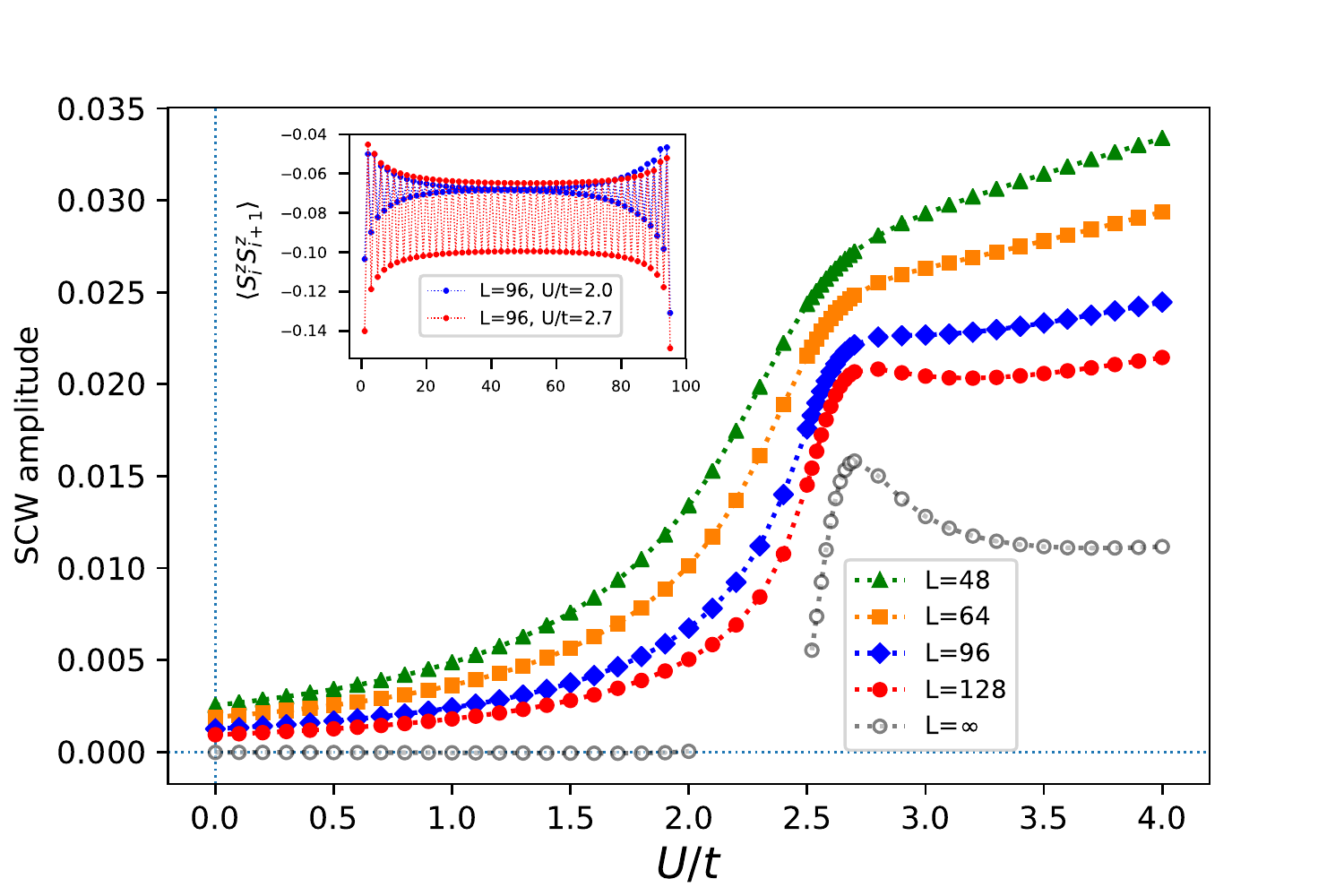}
		\par\end{centering}
	\caption{Spin dimerization order parameter $SCW$ for finite length chains with $L$ ranging from 48 to 128 sites. 
		Extrapolation to the thermodynamic limit is only suggested (hollow circles) where the scaling tendency 
		is well defined. Spin dimerization is absent in the BI phase but appears within the gapless phase, with a peak amplitude roughly where the correlated charge gap opens. 
		Inset: profiles of local correlations $\langle S^z_i S^z_{i+1}\rangle$ for values of $U$ in the BI phase and at the metal-CI transition,
		both for $L=96$ sites chains.
	}
	\label{fig: SCW}
\end{figure}
In order to detect spin dimerization in the ground state we define a n.n.\ spin correlation wave (SCW) order parameter
\begin{equation}
	SCW = -\frac{1}{L-1} \sum_{i=1}^{L-1} (-1)^{i} \langle S^z_iS^z_{i+1} \rangle .
	\label{scw}
\end{equation}
The use of OBC conditions in finite chains explicitly breaks the reflection symmetry
and induces oscillations of the  n.n.\ spin correlations.
In analogy with the discussion of the BOW order, 
we have followed a scaling analysis to separate bulk from boundary contributions.
It suggests a clear thermodynamic limit for low and high values of $U$ 
but does not provide a well defined scaling tendency in the intermediate region. 
Our finite size results and the suggested extrapolation, where confident,  
are shown in Fig.\ \ref{fig: SCW}; the inset illustrates the presence (or absence) of the SCW in the bulk.
The results support that the spin dimerization takes place within the gapless phase, 
with a peak amplitude  where the correlated charge gap opens.
By comparing with Fig. \ref{fig: BOW} it is apparent that the BOW order and the spin dimerization belong together.

\begin{figure}
	\begin{centering}
		\includegraphics[scale=0.55]{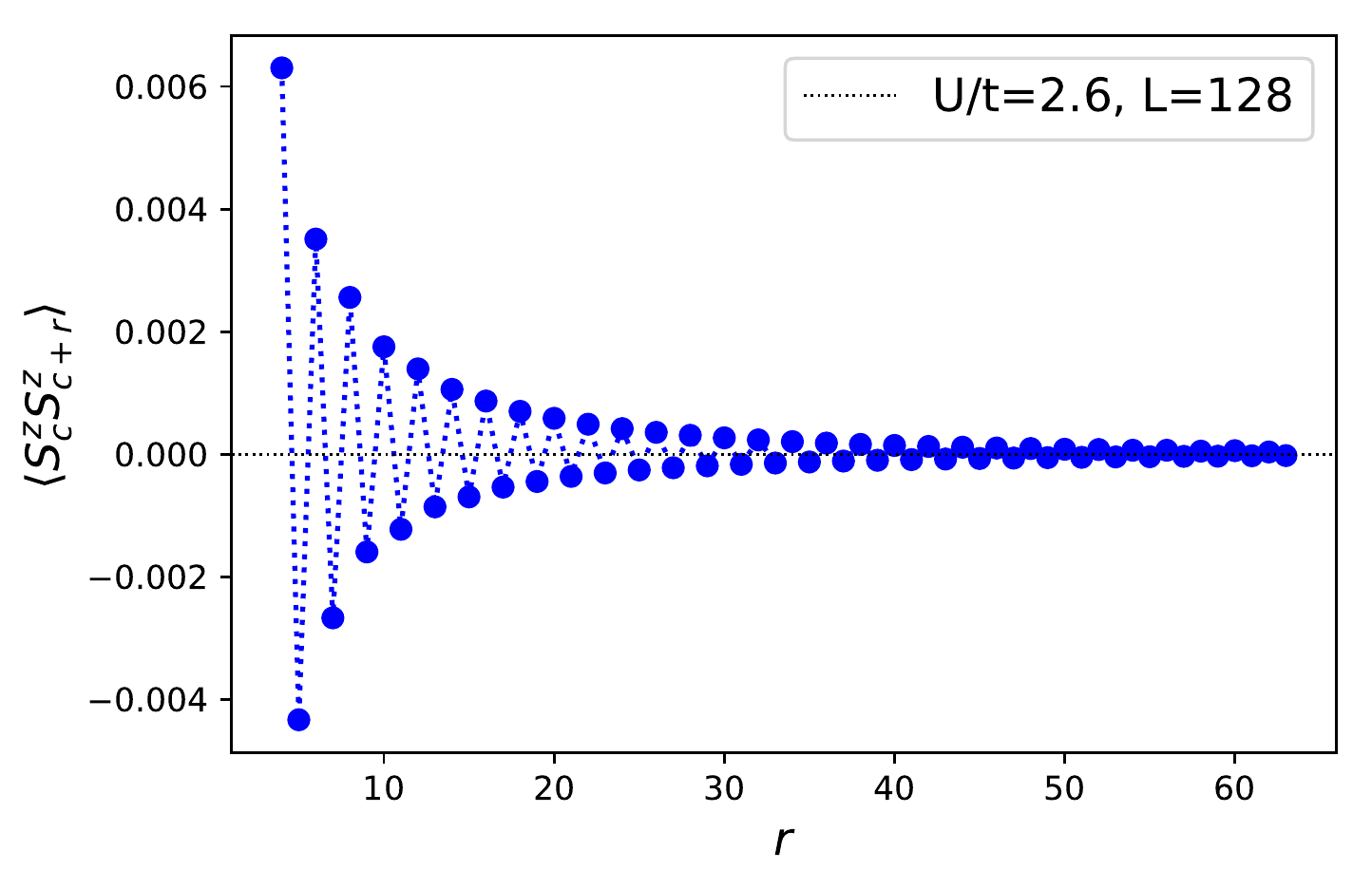}
		\par\end{centering}
	\caption{
		Large distance spin correlations $\langle S^z_{c} S^z_{c+r}\rangle$ where $c=L/2$ is 
		a central site in a $L=128$ sites chain, for $U=2.6$ in the metallic region. 
		An antiferromagnetic order is present, but it is hard to distinguish 
		whether correlations follow an exponential decay with large correlation length 
		or an inverse distance power law.
		}
	\label{fig:SzSz_decay}
\end{figure}

Farther neighbors spin-spin correlations decay with distance. 
The observed decay rate is compatible with an exponential behavior in the BI phase, 
with a correlation length of a few sites that increases as the spin gap decreases with larger U. 
In the CI phase our data is compatible with a quasi-long range antiferromagnetic order, 
with alternate correlations decaying 
like an inverse distance  
power law; this is 
consistent with the mapping into a $J-J'$ Heisenberg spin chain discussed in Section \ref{sec:bosonization} 
and the very small spin gap discussed in Section \ref{subsec:Energy gaps}.
In the intermediate metallic region we observe the formation of a short range antiferromagnetic order, 
as illustrated in Fig.\ \ref{fig:SzSz_decay} in a chain of $L=128$ sites for $U=2.6$. 
The decay rate presumably undergoes a crossover from exponential, with a large correlation length, 
into a quasi-long range order. 
 
\section{Summary and conclusions\label{sec:Summary}}

In the present work we investigate the ground state of an extended one dimensional ionic Hubbard 
model with nearest neighbors hopping $t$, next-to-nearest neighbors hopping $t'$, ionic potential 
$\Delta$ and Hubbard on-site repulsion $U$,  setting  $t'$ in an intermediate regime
where previous studies \cite{Sekania_etal_22} have not been conclusive. We restrict the analysis to half-filling and zero 
magnetization states.

We have focused on a fixed value of ${t^{\prime}}$ and $\Delta$,
where the free $t-t'-\Delta$ chain is still an indirect gap insulator, 
close to the would-be Lifshitz transition if $\Delta$ was absent.
Then we investigate the effects of the Hubbard repulsion. 
Numerically, we set ${t^{\prime}=0.55\,t}$ and $\Delta=0.8\,t$.
Because for the selected set of model parameters the low energy physics of 
the non-interacting particles is given by excitations with non-linear dispersion, 
it is a challenge to analyze the effect of electron-electron interactions $U$ on the system.

\begin{figure}
	\begin{centering}
		\includegraphics[scale=0.55]{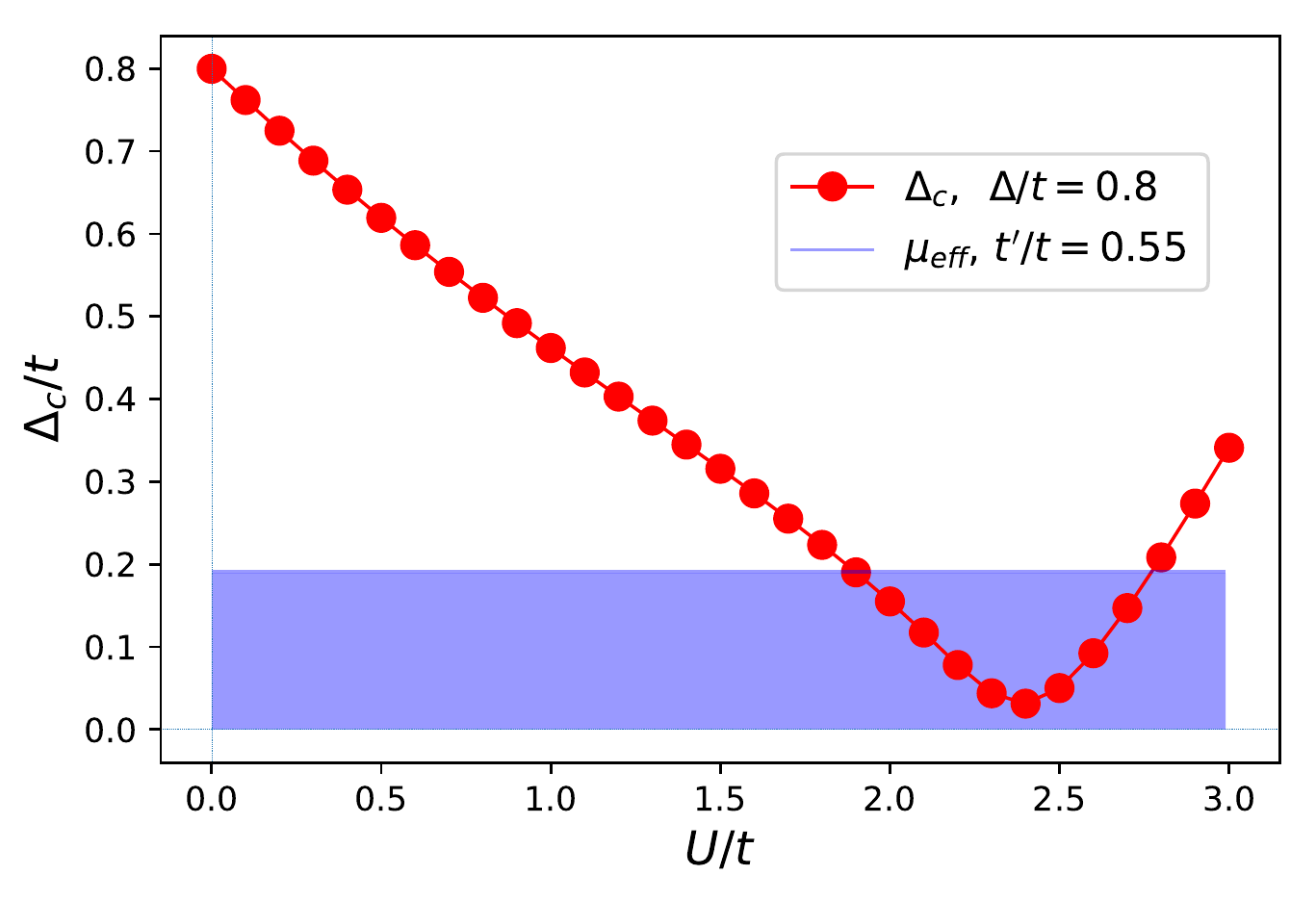}
		\par\end{centering}
	\caption{Qualitative argument comparing the charge gap for the standard IHM ($\Delta=0.8\,t$, $t'=0$, rough DMRG computation in red circles) 
		and the chemical potential introduced by $t'>0.5\,t$ in the bosonization approach (for $t'=0.55\, t$). 
		When the $t'=0$ charge gap is lower than the effective Fermi level (blue surface), fluctuations dominate and the metallic state is stabilized. 
		Spontaneous generation of the BOW order occur inside the metallic phase, making it highly unusual.
		The estimated boundaries of the insulator phases in Fig.\ \ref{fig: 2p energy gap} are compatible with this simple picture. 
	}
	\label{fig:argument}
\end{figure}

On the analytical side we have followed a bosonization approach starting from the free fermion system, with two commensurate Fermi momenta. 
As $t'>0.5\,t$ shifts the Fermi points, a chemical potential is introduced to reestablish them so that perturbations due to 
$\Delta$, $U$ and $t'$ can be treated on equal foots. 
It comes out that three independent length-scales determine the behavior of the ground state: 
one associated to the renormalized ionic gap, one associated to the Hubbard correlated gap and a third one associated to the chemical potential.
When the chemical potential exceeds the ionic and correlated gaps, the metallic phase is established 
by means of a commensurate-incommensurate transition.
Features of the standard IHM, such as the appearance of the BOW order and dominance of correlations, 
occur within this metallic phase while the charge gap remains zero. 
Instead, when the ionic gap or the correlated gap (excluding each other) become  larger than the effective chemical potential, 
the band insulator or the correlated insulator phases, respectively, are formed. 
This findings can be qualitatively appreciated in Fig.\ \ref{fig:argument} where we compare 
the charge gap obtained for the IHM ($t'=0$) with the chemical potential due to $t'>0.5\,t$.


On the numerical side we have explored a wide range of $U$ using the DMRG technique. 
We show that the Hubbard repulsion competes with the ionic free electron state, 
reducing the charge gap. 
Though a vanishing gap makes it difficult to separate the ground state from excitations,
our finite size results suggest that the Hubbard repulsion
drives the system into a gapless ground state at some critical point $U_{c,1}$. 
 This is reminiscent of the so called interaction-resistant metals  \cite{Capone-2021}.
The gapless state is alleged to persist in a wide window of $U_{c,1} < U < U_{c,2}$, 
with neither charge gap nor spin gap and with a long range order CDW pattern induced by the ionic potential. 
After a critical point $U_{c}^\ast$  ({$U_{c,1} < U_{c}^\ast < U_{c,2}$)} the state also supports
short range antiferromagnetic order, 
spontaneous charge bond order and  nearest neighbors spin correlation dimerization.
 This features characterize a very unusual metallic state.

Larger repulsion gives rise to a correlated insulator (Mott-like) phase at some critical point $U_{c,2}$.
The charge gap opens linearly with $U$, 
while the spin gap also opens slightly after  $U_{c,2}$ showing a small peak 
to decay later presumably not closing at any $U$. 
The CDW and the BOW, with decaying amplitude, 
coexist with quasi-long range antiferromagnetic order in this correlated insulator phase .

Additional analytical insight is obtained for large $U$ by freezing the charge degrees of freedom at one electron per site, 
thus mapping the model into a spin $S=1/2$ Heisenberg $J-J'$ chain. As one gets $J'/J>1/4$, the spin system lays in the dimerized phase. This explains the persistence of the BOW order and the finite $1/U$ spin gap within the explored range of $U$.

We expect that the present predictions could be traced in fermionic cold atoms systems in suitable engineered optical lattices.

\section*{Acknowledgments}

G.I.J. acknowledges A.A. Nersesyan, M. Sekania and S. Garuchava  for many useful discussions. 
This work was partially supported by 
CONICET (Grant No. PIP 2021-1146), Argentina, 
and the Shota Rustaveli Georgian National Science Foundation through
the grant FR-19-11872.


\appendix  

\section{Diagonalization of the ionic chain \label{app: A}}

In this Appendix we consider the exactly solvable case of the $t-t'$ ionic chain given by the Hamiltonian 
\begin{eqnarray}\label{eq:App_t1t2_IH_chain}
 {\cal H}_{t-t'-\Delta_r} &=&     -t\sum_{i,\sigma}^{L} \left(c^{\dagger}_{i,\sigma} c\PHDG_{i+1,\sigma}+
  \mathrm{H.c.}\right)\nonumber\\  
 &+&t^{\prime}\sum_{i,\sigma}^{L} \left(c^{\dagger}_{i,\sigma} c\PHDG_{i+2,\sigma}+ \mathrm{H.c.}\right)\\
& + &\frac{\Delta_{r}}{2} \sum_{i,\sigma}^{L}(-1)^{i}n_{i,\sigma}  \, .\nonumber
\end{eqnarray} 
To diagonalize the Hamiltonian (\ref{eq:App_t1t2_IH_chain}) it is convenient to introduce a unit cell with two sites and operators
\begin{equation}
  a_{m,\sigma} \equiv c_{2m-1,\sigma},  \quad  b_{m,\sigma} \equiv c_{2m,\sigma},  \quad
  m=1,...,L/2\, \nonumber
\end{equation}
and rewrite the reduced version of the  Hamiltonian in the following way 
\begin{eqnarray}
 \label{eq:App_ham2}
   {\cal H}_{t-t'-\Delta_r} &=& -t\sum_{m,\sigma}\left[a_{m,\sigma}^\dag \left(b\PHDG_{m,\sigma} + b\PHDG_{m-1,\sigma}\right) 
        + \mathrm{H.c.} \right]\nonumber\\
  &+& t^\prime \sum_{m,\sigma}\left[a_{m,\sigma}^\dag a\PHDG_{m+1,\sigma} +
        b_{m,\sigma}^\dag b_{m+1,\sigma}^{\phantom{}} +\mathrm{H.c.}\right]\nonumber\\
  &-& 
  \frac{\Delta_{r}}{2}
    \sum_{m,\sigma}
      \left(
        n^{(a)}_{m,\sigma}
        -
        n^{(b)}_{m,\sigma}
      \right)  \,,
\end{eqnarray}
where ${n^{(a)}_{m,\sigma}=a_{m,\sigma}^\dag a_{m,\sigma}^{\phantom{}},\, 
	n^{(b)}_{m,\sigma}=b_{m,\sigma}^\dag b_{m,\sigma}^{\phantom{}}}$ 
are spin $\sigma$ particle density operators on odd ($a$) and even ($b$) sites, respectively. 

Performing the Fourier transformation 
\begin{align}
 \label{eq:Four_ab}
 \begin{split}
  a_{m,\sigma} &= \sqrt{\frac{2}{L}} \sum_k e^{i k m} a_{k,\sigma}\,,
  \\
  b_{m,\sigma} &= \sqrt{\frac{2}{L}} \sum_k e^{i k (m + \frac{1}{2})} b_{k,\sigma}\,,
 \end{split}
\end{align}
where ${k=\frac{4\pi}{L} \nu}$, 
with integer $\nu$, ${-\frac{L}{4} < \nu \leqslant \frac{L}{4}}$ and introducing a two-spinor
\begin{equation}\label{eq:Spinor}
\Psi^{\dagger}=\begin{pmatrix}
      a^\dagger_{k,\sigma}, b^\dagger_{k,\sigma},
       \end{pmatrix}  \quad \Psi=\begin{pmatrix}
      a^{\phantom{\dagger}}_{k,\sigma}\\
      b^{\phantom{\dagger}}_{k,\sigma}
    \end{pmatrix}
\end{equation}
we rewrite the Hamiltonian  in momentum space as
\begin{equation}\label{eq:Ham_MF}
 {\cal H}_{t-t'-\Delta_r}=\Psi^{\dagger}\hat{\cal H} \Psi 
\end{equation}
where
\begin{equation}\label{eq:Ham_MF_R}
\hat{\cal H}=  \varepsilon^\prime_k\mathbb{I}+ \varepsilon\PHDG_k\hat{\mathbb{\tau}}_{x} - \frac{1}{2}\Delta_{r}\hat{\mathbb{\tau}}_{z}\, ,
\end{equation}
\begin{align}
  \varepsilon\PHDG_k = -2t \cos \frac{k}{2}\,, \quad 
  \varepsilon^\prime_k = \hphantom{-} 2t^\prime \cos k\,, 
\end{align}
$\mathbb{I}$ is an identity matrix and ${\hat{\mathbb{\tau}}_{x}}$ and ${\hat{\mathbb{\tau}}_{z}}$ are Pauli matrices.
Diagonalization of the Hamiltonian in the form (\ref{eq:Ham_MF_R}) is straightforward. The Bogolyubov transformation
\begin{align}
 \label{eq:bog}
 \begin{split}
  a\PHDG_{k,\sigma}
  &=
  \hphantom{-}
  \cos \varphi\PHDG_{k}
  \alpha\PHDG_{k,\sigma}
  +
  \sin \varphi\PHDG_{k}
  \beta\PHDG_{k,\sigma}
  \,,
  \\
  b\PHDG_{k,\sigma}
  &=
  -
  \sin \varphi\PHDG_{k}
  \alpha\PHDG_{k,\sigma}
  +
  \cos \varphi\PHDG_{k}
  \beta\PHDG_{k,\sigma}
  \,,
 \end{split}
\end{align}
where the angles $\varphi_{k,\sigma}$ are chosen as
\begin{align}
 \label{eq:angles}
  \tan 2\varphi\PHDG_{k}
  =
  \frac{2\varepsilon\PHDG_k}{\Delta_{r}}\,,
  \quad 
  \cos 2\varphi\PHDG_{k,\sigma}
  =
  \frac{\Delta_{r}}{\sqrt{4\varepsilon_{k}^2+\Delta^2_{r}}}
  \,,
\end{align}
diagonalizes the Hamiltonian (\ref{eq:Ham_MF}) as
\begin{equation}
 {\cal H}_{t-t'-\Delta_r}
  =
  \sum_{k,\s}
    \left(
      E_{k}^{-}  \alpha^\dagger_{k,\sigma}  \alpha\PHDG_{k,\sigma}
      +
      E_{k}^{+} \beta^\dagger_{k,\sigma} \beta\PHDG_{k,\sigma}   
    \right)
  \,, 
\end{equation}
where
\begin{equation}
 \label{eq:Ek1D}
  E_{k}^{\pm}
  =
  \varepsilon^\prime_{k}
  \pm
  \sqrt{\varepsilon_{k}^2 + (\Delta_{r}/2)^2}
  \,
\end{equation}
are the energy dispersions for $\alpha$- and $\beta$-quasiparticles, respectively.

In the ground state of the half-filled system the $L$ lowest energy states are filled and the rest are empty.
For ${t^{\prime}=0}$, ${E^-_k}$ and ${E^+_k}$ do not overlap and are separated with a direct gap equal to ${\Delta_{r}}$;
all states in the "lower" band are occupied whereas in the "upper" band all states are empty; 
the system is in the insulating state. 
In the case of a finite ${t^{\prime}}$, however, these bands might overlap, 
due to a $k$-dependent energy shift, 
$\varepsilon^\prime_k$.
For ${t,t^\prime>0}$ the global minimum of the upper $E^+_k$ band is always at ${k=\pi}$, 
\begin{equation}
 \label{eq:E+MIN}
  E^+_{k=\pi}
  =
  -2t^\prime
  +
  {|\Delta_{r}|}/{2}
  \,.
\end{equation}
The $E^-_k$ (lower) band shows a richer composition of maxima: at
\begin{equation}
 \label{eq:t_prime_start}
  t^\prime_\ast=0.5t \sqrt{ 1 + (\Delta_{r}/4t)^{2} }- |\Delta_{r}|/8 \,,
\end{equation}  
the position of the global maximum of the lower band is changed from ${k=\pi}$,  
${E^-_{k=\pi} = -2t^\prime - |\Delta_{r}/2|}$ (${t^\prime < t^\prime_\ast}$), 
to ${k=0}$, ${E^-_{k=0} = 2t^\prime - 2t \sqrt{1 + (\Delta_{r}/4t)^{2}}}$ (${t^\prime > t^\prime_\ast}$).
These possibilities are illustrated in Fig.\ \ref{fig: t1-t2 energy gaps_D} in the main text.

Hence, for ${t^\prime < t^\prime_\ast}$, the system is a band insulator with a direct gap in the excitation spectrum
\begin{equation}
 \label{eq:D_direct} 
  \Delta_{\mathrm{dir}}  =   E^+_{k=\pi}  -   E^-_{k=\pi}  =  |\Delta_{r}| \,,
\end{equation}
while for ${ t^\prime_{\ast}<t^{\prime}<t^{\prime}_{c}}$, where
\begin{equation}
  t^\prime_{c} = 0.5t\sqrt{1 + (\Delta_{r}/4t)^{2}} + |\Delta_{r}|/8 
  \label{tc}
\end{equation}
is an insulator with the indirect gap 
\begin{align}
 \label{eq:ExcitaGap2}
  \Delta_{\mathrm{ind}}  &=  | \Delta_{r}|/2  +  2t \sqrt{1 + (\Delta_{r}/4t)^{2}}  -  4t^\prime 
\end{align}
in the excitation spectrum. The gap decreases linearly with increasing $t^\prime$ and vanishes at ${t^{\prime}=t^{\prime}_{c}}$. 
It is useful to reverse the problem and determine the critical value of the effective ionicity parameter 
$\Delta_r^{cr} \geq 0$ corresponding to the metal-insulator transition at given values of the 
parameters $t$ and  $t^\prime$,
\begin{equation}\label{eq:App_Delta_c}
  \Delta_{r}^{cr} =
  \left\{
    \begin{array}{ccc}
      4 t^\prime - t^{2}/t^\prime & \text{for} & t^\prime \geqslant 0.5 t \\[0.5em]
      0                           & \text{otherwise  }
    \end{array}
  \right.
  \,.
\end{equation}
For ${|\Delta_{r}| > \Delta_{r}^{cr}}$ (${|\Delta_{r}| < \Delta_{r}^{cr}}$), 
the system is in an insulating (metalic) state.
Note that for ${t^\prime < 0.5t}$ the system is in insulating phase for any finite value of $|\Delta_{r}|$.

We complete our analysis by evaluating the ground-state charge distribution
in the insulating phase. The average on-site charge density is 
\begin{equation}
\langle n_{i}\rangle=1-(-1)^{i}\delta\rho_{0}\label{eq:App_n_CDW}
\end{equation}
where 
\begin{align}\label{eq:App_delta_n_CDW}
\delta\rho_{0} & =\frac{1}{L}\sum_{i,\s}\left[\langle n_{m,\sigma}^{(a)}\rangle-\langle n_{m,\sigma}^{(b)}\rangle  \,\right]\nonumber \\
 & =\frac{1}{L}\sum_{k,\s}\cos2\varphi\PHDG_{k}\left[\langle\alpha_{k,\sigma}^{\dagger}\alpha\PHDG_{k,\sigma}\rangle -\langle\beta_{k,\sigma}^{\dagger}\beta\PHDG_{k,\sigma}\rangle \right]
\end{align}
is the  charge imbalance between ``$a$'' (odd) and ``$b$'' (even)
sub-lattices (that is the amplitude of the CDW pattern), induced by the ionic $\Delta_{r}$ term.

In the band insulating ground state 
${\langle\alpha_{k,\sigma}^{\dagger}\alpha\PHDG_{k,\sigma}\rangle=1}$
	and
	${\langle\beta_{k,\sigma}^{\dagger}\beta\PHDG_{k,\sigma}\rangle=0}$ 
for ${-\pi<k\leq\pi}$. Therefore 
\begin{eqnarray}
\delta\rho_{0} & = & \frac{1}{2\pi}\int_{0}^{\pi}\!\!dk\,\cos2\varphi\PHDG_{k}=
\frac{\Delta_{r}\kappa{K}(\kappa)}{2\pi t}\,,\label{eq:delta_rho}
\end{eqnarray}
where $K(\kappa)$ is the complete elliptic integral of the first kind with the modulus 
\begin{equation}
\kappa=\left[1+\left({\Delta_{r}}/{4t}\right)^{2}\,\right]^{-\frac{1}{2}}\,.\label{eq:kappa}
\end{equation}

\section{The band insulator phase \label{app: B}}


To assess the accuracy of the 2-FP approach let us apply the  bosonization analysis in the exactly
solvable case of the free $t-t'$ ionic chain, where the Hubbard repulsion is included only via the renormalization of the 
ionic term.  At $U=0$  the system is decoupled into the identical 
"up" and "down" spin component parts 
${\cal H} = \int  dx \left[ h_{\up} + h_{\down}\right]$, 
where for each spin component the Hamiltonian is the sine-Gordon model with topological term 
\bea
h_{\s} &=& \frac{v_{F}}{2}\big[
\left(\partial_{x}\phi_{\s}\right)^{2} +
\left(\partial_{x}\theta_{\s}\right)^{2}\big]- \frac{\mu_{\text{eff}}} {\sqrt{\pi}}\partial_{x}\phi_{\s}\nonumber\\
&-&\frac{\Delta_{r}}{2\pi \alpha_{0}}\sin\sqrt{4\pi}\phi_{\s}\, ,  \quad(\s =\up , \down) \label{H-sigma_BOS}
\eea
with $\mu_{\text{eff}}$ given by Eq.\ (\ref{ChemPot_eff}) in the main text.
Each of these Hamiltonians is the standard one for the commensurate-incommensurate transition~\cite{JN_78,PT_79}. 
At $\mu_{{\it eff}}=0$, the model is described by the theory of two commuting sine-Gordon fields with
$\beta^{2}=4\pi$. In this case the excitation spectrum is gapped and
the excitation gap is given by the mass of the "up" ("down") field
soliton $M_{\up}=M_{\down}=\Delta_{r}/2$. In the ground state the
$\phi_{\up}$ and $\phi_{\down}$ fields are pinned with vacuum
expectation values $\langle 0|\phi_{\s}|0 \rangle = \sqrt{\pi}(n +
1/4)$ with integer $n$ what gives the LRO  in-phase distribution of electron density in the ground state
\bea \rho_{c}(x) &\simeq&  (-1)^{n}\frac{1}{\pi
\alpha_{0}}\sum_{\s}\sin(\sqrt{4\pi}\phi_{\s}(x)).\nonumber\\
 \label{upDown_density} \eea
Thus at low $t^{\prime}<0.5t$ ($\mu_{\text{eff}} = 0$) the ground state of the system corresponds to 
a CDW type band insulator with a single energy scale given by the ionic potential $\Delta_{r}$.

At  $t^{\prime}>0.5t$ ($\mu_{\text{eff}} \neq 0$) it is necessary to consider the ground state
of the sine-Gordon model in sectors with nonzero topological charge.
Competition between the chemical potential term ({\em i.e.} ${t^{\prime}>0.5t}$)
and the commensurability energy given by ${\Delta_{r}}$ finally drives a
continuous phase transition from a gapped (insulating) phase at
$\mu_{\text{eff}} < \mu_{\text{eff}}^{c}$ to a gapless (metallic)
phase at $\mu_{\text{eff}} > \mu_{\text{eff}}^{c}$, where
\be  \mu_{\text{eff}}^{c} = \Delta_{r}/2\, .
\label{Critical_mu}
\ee
Using Eq.\ (\ref{ChemPot_eff}) we easily obtain that the critical value
of the n.n.n.\ hopping amplitude $t^{\prime}$, 
obtained  in the 2-FP approach from the condition
(\ref{Critical_mu}), coincides with the exact value 
given in (\ref{tc}).

As we observe, 
the insulator-metal transition at $t^{\prime}>t^{\prime}_{c}$ is connected with a change of the topology of
the Fermi surface and a corresponding redistribution of the electrons
from the lower ("-") band into the upper ("+") band. 
At the transition point the derivative of the ground state energy with respect to the chemical potential displays 
a singular behavior of the usual square-root type  $\partial E_{0}/\partial \mu \sim -(\mu -\mu_{c})^{1/2}$ \, 
when the chemical potential is constant, or linear dependence $\partial E_{0}/\partial \mu \sim -(\mu -\mu_{c})$  
in the case of fixed particle density~\cite{Vekua_2009}.


\end{document}